\documentclass[useAMS,usenatbib]{mn2e}
\usepackage{myaasmacros}
\usepackage{graphicx}
\usepackage{ulem}
\usepackage{amsmath}
\usepackage{amssymb}
\usepackage[table]{xcolor}

\def\ltsima{$\; \buildrel < \over \sim \;$}
\def\simlt{\lower.5ex\hbox{\ltsima}}   
\def\gtsima{$\; \buildrel > \over \sim \;$}
\def\simgt{\lower.5ex\hbox{\gtsima}}

\newcommand{\pc} {{\,\rm pc}}

\def\Msun{\, M_{\odot}}

\usepackage[pdftex,
        colorlinks=true,
        urlcolor=blue,		
        filecolor=blue,		
        citecolor=blue,		
        linkcolor=blue,		
        pdftitle={Untitled},
        pdfauthor={},
        pdfsubject={},
        pdfkeywords={},
        pdfproducer={pdfLaTeX},
        pagebackref,
        pdfpagemode=None,
        bookmarksopen=true]{hyperref}

\def\coreNFW{{\sc coreNFW}}
\def\bfcoreNFW{{\sc \bf coreNFW}}
\def\EMCEE{{\sc emcee}}
\def\Barolo{{\sc $^{3\rm D}$Barolo}}
\def\SFRN{{\rm SFR}_{0.1\,{\rm Gyr}} / \langle {\rm SFR} \rangle}

\usepackage{array}
\newcolumntype{L}[1]{>{\raggedright\let\newline\\\arraybackslash\hspace{0pt}}m{#1}}
\newcolumntype{C}[1]{>{\centering\let\newline\\\arraybackslash\hspace{0pt}}m{#1}}
\newcolumntype{R}[1]{>{\raggedleft\let\newline\\\arraybackslash\hspace{0pt}}m{#1}}

\title[Dwarf rotation curve shape \& diversity in $\Lambda$CDM]{Understanding the shape and diversity of dwarf galaxy rotation curves in $\Lambda$CDM}

\author[J. I. Read et al.]{J. I. Read$^{1}$\thanks{E-mail: justin.inglis.read@gmail.com}, G. Iorio$^{2,3}$, O. Agertz$^1$, F. Fraternali$^{2,4}$\\
$^1${\small Department of Physics, University of Surrey, Guildford, GU2 7XH, Surrey, UK}\\
$^2${\small Dipartimento di Fisica e Astronomia, Universit\`a di Bologna, Viale Berti Pichat 6/2, I-40127, Bologna, Italy}\\
$^3${\small INAF -- Osservatorio Astronomico di Bologna, via Ranzani 1, I-40127, Bologna, Italy}\\
$^4${\small Kapteyn Astronomical Institute, University of Groningen, Landleven 12, 9747 AD Groningen, The Netherlands}\\
}

\begin{document}

\maketitle

\begin{abstract}
The shape and diversity of dwarf galaxy rotation curves is at apparent odds with dark matter halos in a $\Lambda$ Cold Dark Matter ($\Lambda$CDM) cosmology. We use mock data from isolated dwarf galaxy simulations to show that this owes to three main effects. Firstly, stellar feedback heats dark matter, leading to a `\coreNFW' dark matter density profile with a slowly rising rotation curve. Secondly, if close to a recent starburst, large HI bubbles push the rotation curve out of equilibrium, deforming the rotation curve shape. Thirdly, when galaxies are viewed near face-on, their best fit inclination is biased high. This can lead to a very shallow rotation curve that falsely implies a large dark matter core. All three problems can be avoided, however, by a combination of improved mass models and a careful selection of target galaxies. Fitting our \coreNFW\ model to mock rotation curve data, we show that we can recover the rotation curve shape, dark matter halo mass $M_{200}$ and concentration parameter $c$ within our quoted uncertainties.

We fit our \coreNFW\ model to real data for four isolated dwarf irregulars, chosen to span a wide range of rotation curve shapes. We obtain an excellent fit for NGC 6822 and WLM, with tight constraints on $M_{200}$, and $c$ consistent with $\Lambda$CDM. However, IC 1613 and DDO 101 give a poor fit. For IC 1613, we show that this owes to disequilibria and its uncertain inclination $i$; for DDO 101, it owes to its uncertain distance $D$. If we assume $i_{\rm IC1613} \sim 15^\circ$ and $D_{\rm DDO101} \sim 12$\,Mpc, consistent with current uncertainties, we are able to fit both galaxies very well. We conclude that $\Lambda$CDM appears to give an excellent match to dwarf galaxy rotation curves.
\end{abstract}


\section{Introduction}\label{sec:intro}
Galaxy rotation curves have provided some of the earliest and most compelling evidence for dark matter in the Universe \citep{1959BAN....14..323V,1970ApJ...159..379R,Freeman1970,1980ApJ...238..471R,1985ApJ...295..305V}. In all galaxies observed to date, rotation curves fall far more slowly than would be expected from the visible light and gas alone. This is typically taken as evidence for an exotic missing mass component -- most likely a new particle that lies beyond the standard model of particle physics\footnotemark\ \citep[e.g.][]{1996PhR...267..195J,2009ARNPS..59..191B,2010pdmo.book.....B,2014JPhG...41f3101R}. However, the nature and properties of such a particle remain unknown. 

\footnotetext{An alternative explanation is to modify weak-field gravity \citep[e.g.][]{1983ApJ...270..365M,2004PhRvD..70h3509B,2006JCAP...03..004M}. However, this runs into difficulties when faced with data from gravitational lensing and/or the growth of large scale structure (e.g. \citealt{2006ApJ...648L.109C,2006MNRAS.368..171Z,2011IJMPD..20.2749D}).}

While rotation curves have long given us evidence for dark matter, there is an enduring puzzle relating to their shape. Almost all rotation curves appear to rise less steeply than is predicted by pure dark matter simulations of structure formation in the standard $\Lambda$CDM cosmological model. Such simulations predict central dark matter density profiles that rise as $\rho \sim r^{-1}$ (a `cusp'; \citealt{1991ApJ...378..496D,1994Natur.370..629M,1996ApJ...462..563N}) in stark contrast with observed rotation curves that favour constant density `cores' \citep[e.g.][]{1994ApJ...427L...1F,2001ApJ...552L..23D,2009arXiv0910.3538D,2011AJ....142...24O,2015AJ....149..180O,2013MNRAS.433.2314H,2014MNRAS.443.3712H,2014ApJ...789...63A,2015AJ....149..180O}. This has become known as the {\it cusp-core problem}. 

One solution to the cusp-core problem is to suggest that there is a problem with the rotation curve data, or with the interpretation of these data \citep[e.g.][]{2000AJ....119.1579V,2003ApJ...583..732S,2004MNRAS.355..794H}. To explore this, \citet{2004ApJ...617.1059R} used simulated mock rotation curves to test how well they trace the underlying gravitational potential. They focussed on three effects: inclination correction; asymmetric drift correction (important if the gas velocity dispersion is a significant fraction of the rotational velocity; see \S\ref{sec:asymcorrect}); and non-circular motions due to a central bar. They found that the rotation curve is typically biased by about $\sim 20$\% due to such effects, though in some extreme cases it can be shifted by up to a factor of two. \citet{2007ApJ...657..773V} revisited this issue for the specific case of claimed dark matter cores in NGC 3109 and NGC 6822 (the latter of which we will consider here in some detail also). Using numerical simulations as mock data, they performed an end-to-end recovery of the underlying dark matter distribution. They found that if they do not properly correct for thermal and turbulent gas pressure support in the disc (the `asymmetric drift' correction; see \S\ref{sec:asymcorrect}), then they falsely favour a dark matter core over the correct solution that is cuspy. However, the simulated discs in their study (with a peak rotation velocity of $\sim 70$\,km/s) were rather hot, with a turbulent gas dispersion of $\sigma_{\rm turb} \sim 22$\,km/s (inflated by streaming motions along a central bar), and a gas sound speed of $c_s \sim 60$\,km/s. NGC 6822, that has a comparable peak rotation velocity, has a measured dispersion of just $\sigma_{\rm gas} \sim 6$\,km/s that is $\sim$ constant across the disc (\citealt{2003MNRAS.340...12W}; and see \S\ref{sec:asymcorrect} for a discussion of how $\sigma_{\rm gas}$ relates to $\sigma_{\rm turb}$ and $c_s$). Other galaxies of similar peak rotation velocity are observed to be similarly cold, with $\sigma_{\rm gas}$ in the range $6 < \sigma_{\rm gas} < 15$\,km/s \citep[e.g.][]{2015AJ....149..180O}. This likely explains why in a later study \citet{2011MNRAS.414.3617K} find -- seemingly at odds with \citet{2007ApJ...657..773V} --  that they are able to successfully disentangle cusps and cores from mock 2D velocity field data, despite ignoring asymmetric drift corrections altogether. Finally, \citet{2011AJ....142...24O} extract mock rotation curves from dwarf galaxy simulations taken from \citet{GovernatoEtAl2010}. They perform and end-to-end analysis similar to that in \citet{2007ApJ...657..773V}, finding that they are also able to correctly recover the underlying dark matter distribution.

If dark matter cores are not simply a misinterpretation of observational data, then this opens the door to more exotic explanations. Many authors have suggested that dark matter cores could point to new physics, for example self-interacting or scalar-field dark matter models \citep[e.g.][]{2000PhRvL..84.3760S,2002CQGra..19.5017A,2013MNRAS.431L..20Z,2014arXiv1412.1477E,2012JPhCS.378a2012M}, or weakly relativistic warm dark matter \citep[e.g.][]{2001ApJ...556...93B,2001ApJ...559..516A,2007PhRvD..75f1303S,2009ARNPS..59..191B,2011JCAP...03..024V,2012MNRAS.424.1105M}. However, before we can conclude that any of these possibilities are favoured by the data, we must first be confident of our model predictions in $\Lambda$CDM. It is important to remember that dark matter cusps are a prediction of {\it pure dark matter} structure formation simulations. Implicit in these simulations is an assumption that baryons -- stars and gas -- have little or no impact on the underlying dark matter distribution. There has been a significant debate in the literature about the validity of this approximation. \citet{1996MNRAS.283L..72N} were the first to propose that a central cusp could be transformed to a core by impulsive gas loss driven by supernova explosions. They found that, for reasonable initial conditions and gas collapse factors, the effect of a single burst is very small \citep[see also][]{2002MNRAS.333..299G}. However, \citet{2005MNRAS.356..107R} showed that {\it multiple} repeated bursts can cause this small effect to accumulate, gradually grinding a dark matter cusp down to a core. Such an effect has now been observed in high resolution hydrodynamic simulations that resolve the interstellar medium (e.g. \citealt{2008Sci...319..174M,GovernatoEtAl2010,2013MNRAS.429.3068T,2014MNRAS.441.2986D,2015MNRAS.446.1140T,2015arXiv150202036O, 2015arXiv150804143R}; and for a review see \citealt{2014Natur.506..171P}). The physics of such cusp-core transformations is now well understood \citep{2012MNRAS.421.3464P,2015arXiv150207356P}, while there is mounting observational evidence for the bursty star formation that is required to drive the process \citep[e.g.][]{2012ApJ...750...33L,2013MNRAS.429.3068T,2012ApJ...744...44W,2014MNRAS.441.2717K,2015MNRAS.450.3886M,2015arXiv150804143R,2015arXiv151201235E}.

The above progress suggests that cusp-core transformations driven by stellar feedback likely explain the observed shallow rise seen in many dwarf galaxy rotation curves. However, recently a new problem has emerged: the {\it diversity} of dwarf galaxy rotation curves. \citet{2015arXiv150401437O} compared dwarfs from recent cosmological hydrodynamic simulations with a wide array of observed dwarf galaxy rotation curves, including those from the THINGS and Little THINGS surveys (\citealt{2008AJ....136.2648D,2011AJ....142...24O,2012AJ....144..134H,2015AJ....149..180O}; we refer the reader to \citealt{2015arXiv150401437O} for a full list of the data they use). They found that the data show a wide variety of rotation curve shapes at fixed peak rotation velocity, in stark contrast with their simulations that show remarkably little scatter. In particular, some galaxies exhibit an extremely shallow rise that appears to imply truly massive dark matter cores $\simgt 4$\,kpc -- far larger than the $\simlt 1$\,kpc cores predicted by recent simulations. \citet{2016arXiv160101026O} suggest that this could owe to systematic inclination and/or distance errors in the rotation curve reconstruction, but it remains to be seen whether such large systematic errors are plausible. By contrast, \citet{2015MNRAS.454.1719B} suggest that the diversity could owe to the expected halo-to-halo scatter in $\Lambda$CDM that is amplified by dark matter cusp-core transformations.

In this paper, we use two recent high resolution simulations of isolated dwarf galaxies (of mass $M_{200} = 5 \times 10^8$\,M$_\odot$ and $10^9$\,M$_\odot$) to shed light on both the shape and diversity of dwarf galaxy rotation curves. Our simulations reach a minimum cell size of $4$\,pc, allowing us to resolve the effect of individual supernova explosions. At this resolution, we become insensitive to our `sub-grid' numerical parameters, making the simulations substantially more predictive \citep[][hereafter R16]{2015arXiv150804143R}. In R16, we showed that our simulated dwarfs give an excellent match to the photometric light profiles; star formation histories; metallicity distribution functions; and star/gas kinematics of low mass isolated dwarfs in the field without any fine-tuning of the model parameters. Here, we use these simulations to create dynamically realistic mock rotation curve data. We start by assuming that we can reliably correct for inclination and asymmetric drift, looking first at the effect of HI bubbles driven by stellar feedback on the rotation curve, and the importance of dark matter cusp-core transformations. We then consider how well we can reconstruct the inclination and asymmetric drift corrected rotation curve from mock inclined HI data cubes. In all cases, we fit our mock rotation curves using the \EMCEE\ python package of \citet{2013PASP..125..306F} to test how well we can recover the halo mass $M_{200}$ and concentration parameter $c$ (see equation \ref{eqn:rhoNFW}). In performing these fits, we make use of our new \coreNFW\ dark matter halo profile that accounts for cusp-core transformations driven by stellar feedback (\S\ref{sec:massmodel}; and R16). Finally, we apply our rotation curve fitting method to real data for four isolated dwarf galaxies: NGC 6822; WLM; IC 1613; and DDO 101, chosen to span a range of interesting rotation curve shapes. Using the insight gained from our mock data analysis, we discuss why we obtain an excellent fit for two of these galaxies (NGC 6822 and WLM) but seemingly not for the other two (IC 1613 and DDO 101).

This paper is organised as follows. In \S\ref{sec:simulations}, we briefly review the numerical simulations (these are discussed in more detail in R16). In \S\ref{sec:data}, we describe our data compilation for NGC 6822, WLM and IC 1613, and we briefly describe the \Barolo\ method for extracting rotation curves from these data \citep{2015MNRAS.451.3021D}. In \S\ref{sec:rotmethod}, we describe our rotation curve fitting method that makes use of our \coreNFW\ profile and the \EMCEE\ code. In \S\ref{sec:mock}, we apply this method to our mock data. In \S\ref{sec:resultsdata}, we apply our method to real data for four isolated dwarf galaxies: NGC 6822; WLM; IC 1613; and DDO 101. Finally, in \S\ref{sec:conclusions} we present our conclusions.

\section{The simulations}\label{sec:simulations}

The simulations are described in detail in R16. Briefly, we set up equilibrium isolated dwarf galaxies following \citet{2006MNRAS.tmp..153R}. The particles were populated using accept/reject from an analytic density profile; their velocities were drawn from a numerically calculated distribution function, assuming an isotropic velocity dispersion tensor. For the initial conditions, we assumed a \citet{1996ApJ...462..563N} (hereafter NFW) dark matter density profile: 

\begin{equation} 
\rho_{\rm NFW}(r) = \rho_0 \left(\frac{r}{r_s}\right)^{-1}\left(1 + \frac{r}{r_s}\right)^{-2}
\label{eqn:rhoNFW}
\end{equation}
where the central density $\rho_0$ and scale length $r_s$ are given by: 
\begin{equation} 
\rho_0 = \rho_{\rm crit} \Delta c^3 g_c / 3 \,\,\,\, ; \,\,\,\, r_s = r_{200} / c;\,\,{\rm with} 
\end{equation}
\begin{equation}
g_c = \frac{1}{{\rm log}\left(1+c\right)-\frac{c}{1+c}}
\label{eqn:gc}
\end{equation}
and
\begin{equation} 
r_{200} = \left[\frac{3}{4} M_{200} \frac{1}{\pi \Delta \rho_{\rm crit}}\right]^{1/3}
\label{eqn:r200}
\end{equation} 
where $c$ is the dimensionless {\it concentration parameter}; $\Delta = 200$ is the over-density parameter; $\rho_{\rm crit} = 136.05$\,M$_\odot$\,kpc$^{-3}$ is the critical density of the Universe at redshift $z=0$; $r_{200}$ is the `virial' radius at which the mean enclosed density is $\Delta \times \rho_{\rm crit}$; and $M_{200}$ is the `virial' mass -- the mass within $r_{200}$.

These halos were then filled with the universal baryon fraction in gas, $f_b = 0.15$, set up also as an NFW profile in hydrostatic equilibrium. The gas was given a seed metallicity of $Z_{\rm gas}=10^{-3}\,Z_\odot$, representing Pop III enrichment \citep[e.g.][]{2001ApJ...548...19N,2009arXiv0908.1254B,2013RvMP...85..809K}. We added angular momentum to the gas assuming a specific angular momentum profile as in \citet{BullockEtal2001b}: 

\begin{equation} 
j(r) \simeq j_{\rm max} \frac{M_{\rm NFW}(<r)}{M_{200}}
\end{equation}
where $M_{\rm NFW}$ is the NFW halo cumulative mass profile that follows from equation \ref{eqn:rhoNFW}; and the peak specific angular momentum $j_{\rm max}$ is set such that the total halo angular momentum is given by: 

\begin{eqnarray} 
J_{\rm 200} & = & 4\pi \int_0^{\infty} j(r) \rho_{\rm NFW}(r) r^2 dr \\
& = & \lambda' \sqrt{2 G M_{200}^3 R_{\rm 200}}
\end{eqnarray}
where $\lambda'$ is the {\it spin parameter}. We assume the cosmic mean value $\lambda' = 0.035$.

We consider here two of the simulations from R16 set up as above with masses $M_{200} = 5\times 10^8$ and $10^9$\,M$_\odot$, labelled M5e8c25\_2e6 and M9c22\_4e6, respectively. They both have a dark matter particle resolution of $m_{\rm dm} = 250$\,M$_\odot$; a finest grid cell size gas resolution of $\Delta x\approx 4\pc$; and a stellar sampling mass of $m_* = 300\Msun$ \citep[see e.g.][]{dubois08}. Note that the virial masses $M_{200}$ refer to the {\it total mass} in gas plus dark matter. The dark matter halo masses for these models are slightly lower: $M_{200,{\rm DM}} = M_{200} (1-f_b) = 0.85 M_{200}$ (R16).

The simulations were evolved for 14\,Gyrs each, using the Adaptive Mesh Refinement (AMR) code {\small RAMSES} \citep{teyssier02} with cooling, star formation and feedback physics prescriptions as described in detail in \cite{2013ApJ...770...25A}; \cite{2015ApJ...804...18A} and R16.

The true gas spatial resolution of the simulations in R16 is somewhat larger than the minimum cell size. However, even using the \citet{1997ApJ...489L.179T} criteria of $\sim 4 \Delta x \approx 16$\,pc, the resolution scale is substantially smaller than the projected half stellar mass radii of the simulated galaxies ($R_{1/2} \sim 200 - 500$\,pc). The simulations are also robust to spurious two-body relaxation over a Hubble time \citep[e.g.][]{2003MNRAS.338...14P} on scales $\simgt 40$\,pc (see R16, Appendix A1).

The key result from these R16 simulations was that star formation over a Hubble time transformed the initial NFW dark matter distribution into a `\coreNFW' profile that has $\sim$ constant density within $R_{1/2}$ (see \S\ref{sec:massmodel} for a complete description of the \coreNFW\ profile). In R16, we showed that this result is robust to order-of-magnitude changes in the numerical `sub-grid' physics parameters and/or initial conditions. As a result, we claimed that the \coreNFW\ profile is the correct prediction for a $\Lambda$CDM cosmology (with baryons) on mass scales $10^8 \simlt M_{200}/{\rm M}_\odot \simlt 10^{10}$, at least to within a factor $\sim 2$ scatter in the \coreNFW\ profile parameters\footnotemark. This robustness is a direct result of the numerical resolution in R16 that has two key effects: (i) we correctly capture the momentum injection into the interstellar medium (ISM) generated by expanding bubbles of shock heated gas \citep[e.g.][]{2015arXiv150105655K}; and (ii) star formation becomes self-regulated by stellar feedback \citep[e.g.][]{2003ApJ...590L...1K,2008PASJ...60..667S,2011MNRAS.417..950H,2013MNRAS.432.2647H,2013ApJ...770...25A}. 

\footnotetext{The simulations in R16 do miss some potentially important physics, for example magnetic fields, radiative transfer, dust and cosmic rays. However, the excellent agreement with a wide range of data for isolated dwarf irregulars, without any fine-tuning of the simulation parameters, suggests that these missing physics are next-to-leading order effects. For further discussion of these points, see R16.}

\subsection{Extracting the rotation curves}

We analyse the $10^9$\,M$_\odot$ simulation at several different times throughout its starburst cycle, focussing on a `quiescent' phase, a `starburst' phase and a `post-starburst' phase (see Figure \ref{fig:mock_common}). We analyse the $5 \times 10^8$\,M$_\odot$ dwarf at a simulation time of 14\,Gyrs, when it is quiescent with little recent star formation (see R16). The simulations were rotated and centred such that the gas discs are aligned with the $x-y$ plane. We then define the gas as being `HI' if it has temperature $T < 10^4$\,K. Under collisional ionisation equilibrium, this is the temperature at which the ionisation fraction is unity. We find that cutting instead on $T = 2 \times 10^4$\,K (where the ionisation fraction is $<10$\%) makes very little difference to our results.

\subsubsection{Asymmetric drift correction}\label{sec:asymcorrect}

We assume that the gas rotational velocity $v_{\phi,{\rm gas}}$ relates to the circular speed: 

\begin{equation} 
v_c^2 = -\frac{1}{R} \left. \frac{\partial \Phi(R,z)}{\partial R}\right|_{z = 0}
\end{equation} 
where $R$ is the cylindrical radius and $\Phi$ is the gravitational potential via an `asymmetric drift' correction:

\begin{equation} 
v_c^2 = v_{\rm rot,gas}^2 = v_{\phi,{\rm gas}}^2 + \sigma_{\rm D}^2
\end{equation}
where $v_{\rm rot,gas}^2$ is the `asymmetric drift corrected rotation curve'; 

\begin{equation} 
\sigma_{\rm D}^2 = -\frac{R}{\Sigma_{\rm gas}} \frac{d}{dR}\left(\Sigma_{\rm gas} \sigma_{\rm gas}^2 \right);
\label{eqn:sigmaD}
\end{equation}
and $\Sigma_{\rm gas}$ and $\sigma_{\rm gas}$ are the gas surface density and gas `effective velocity dispersion', respectively \citep[e.g.][]{1987gady.book.....B}. The gas velocity dispersion $\sigma_{\rm gas}$ includes contributions from both thermal pressure and turbulence \citep[e.g.][]{2007ApJ...657..773V}:   

\begin{equation}
\sigma_{\rm gas}^2=c_{\rm s}^2+\sigma_{\rm turb}^2
\end{equation} 
where $c_{\rm s}$ is the sound speed of the gas and $\sigma_{\rm turb}$ is the gas turbulent velocity dispersion. In the simulated dwarf galaxies investigated in this work, the gas surface density $\Sigma_{\rm gas}$ is well fit by an exponential, though depending on its phase through the starburst cycle it can show prominent wiggles and even an inner hole. For this reason, we do not numerically solve equation \ref{eqn:sigmaD}, but rather use the best-fitting exponential for $\Sigma_{\rm gas}$. In addition, we find that, $\sigma_{\rm turb} \sim 5$\,km/s and $c_{\rm s} \sim 5$\,km/s are nearly constant out to $R \sim 2$\,kpc (independent of the starburst cycle phase), in good agreement with observational data for isolated dwarfs of similar peak rotation velocity \citep[see e.g.][and the discussion in R16]{2015AJ....149..180O}. Thus, to a very good approximation, equation \ref{eqn:sigmaD} simplifies to: 

\begin{equation}
\sigma_{\rm D}^2 \simeq \frac{R}{R_{\rm gas}} \sigma_{\rm gas}^2
\label{eqn:sigmaDsimple} 
\end{equation} 
where $R_{\rm gas}$ is the exponential gas disc scale length.

We discuss more sophisticated asymmetric drift corrections in Appendix \ref{app:asymcorrect} where we explore the effect of a wide range of $\Sigma_{\rm gas}$ profiles on WLM's rotation curve. The differences in both the rotation curve and its implied $M_{200}$ and $c$ lie in every case within our quoted 1-$\sigma$ uncertainties.

We added Gaussian velocity errors of fixed variance $\sigma_{\rm gas,err} = 1$\,km/s to the above rotation curves.

\subsubsection{Mock HI datacubes}\label{sec:mockcubes}

In \S\ref{sec:datacubes}, we consider how well we can reconstruct the rotation curve from mock inclined HI data cubes. We generate these by slicing the data in velocity channels of width $1$\,km/s at a spatial resolution of 50\,pc for a range of inclination angles: $i = 15^\circ, 30^\circ, 45^\circ, 60^\circ, 75^\circ$. We do not explicitly add any broadening due to thermal or turbulent pressure support as our primary goal with these datacubes is to test how well we can recover inclination. To simulate real observations, we added random Gaussian noise to these mock data and then applied both a 2D spatial and a 1D velocity (Hanning) smoothing. In this way, we obtained a simulated observation of the mock datacubes with an instrumental beam of 20 arcsec and with a noise per channel of $\sim 0.027 {\rm M}_\odot / {\rm pc}^2$, that is the typical value found in the Little THINGS datacubes at this resolution \citep{2012AJ....144..134H}.

Our method for extracting rotation curves from these HI data cubes is described in \S\ref{sec:barolo}. Once extracted, we applied an asymmteric drift correction as in \S\ref{sec:asymcorrect} to these rotation curves.

\section{The data}\label{sec:data}

We study four dwarf galaxies with excellent literature data, chosen to span a wide range of rotation curve shapes: NGC 6822; WLM; IC 1613; and DDO 101. The data are summarised in Table \ref{tab:data}. We briefly discuss each galaxy in turn, next. Our method for extracting rotation curves from HI data cubes is described in \S\ref{sec:barolo}.

\vspace{2mm}
\noindent
{\bf NGC 6822: }
NGC 6822 was first discovered by \citet{1884AN....110..125B}. It is one of the closest isolated dwarf irregulars known, lying some $D = 490 \pm 40$\,kpc from the Milky Way \citep{1998ARA&A..36..435M}. For this reason, it has a wealth of excellent data. In particular, its high resolution rotation curve extends to an impressive $\sim 5$\,kpc from the galactic centre \citep{2003MNRAS.340...12W}. It has a relatively smooth HI gas distribution, with the exception of a large HI hole of size 1.4 - 2\,kpc (\citealt{2000ApJ...537L..95D}, and see Figure \ref{fig:real_rotcurve} upper left panel). Several theories have been put forward for the origin of this and similar holes in dwarf irregular galaxies. One possibility is that the hole formed as a result of a merger/interaction with a high velocity gas cloud or smaller companion galaxy \citep[e.g.][]{1987A&A...179..219T,2006ApJ...637L..97B}; another is that the hole results from gravitational instability in the disc (e.g. \citealt{2000ApJ...540..797W}, and for a review see \citealt{1988ARA&A..26..145T}). However, it is now widely accepted that most of these holes, including the one in NGC 6822, owe to stellar feedback \citep{1980ApJ...238L..27B,2001ApJ...563..867S,2012ApJ...747..122C,2012ApJ...746...10K}. \citet{2012ApJ...747..122C} show that there is sufficient energy from star formation to create the hole in NGC 6822 and they estimate an age of $> 500$\,Myrs, consistent with its low/zero expansion velocity. Interestingly, the hole coincides with a notable dip in the rotation curve (see Figure \ref{fig:real_rotcurve}, upper row), a correspondence that we discuss further in \S\ref{sec:datacubes}. However, the rotation curve within $\sim 2.5$\,kpc is also affected by the possible presence of a stellar bar or a misaligned stellar component \citep{2006ApJ...636L..85D} making such a correspondence possibly coincidental. NGC 6822 appears to be at a relatively quiescent moment in its history, having formed stars over the past 0.1\,Gyr at about half of its mean rate over a Hubble time: $\SFRN = 0.51 \pm 0.14$ \citep{2012AJ....143...47Z}.

\vspace{2mm}
\noindent
{\bf WLM: } 
WLM was first discovered by \citet{1909AN....183..187W} and later rediscovered by Lundmark and \citet{1926MNRAS..86..636M} -- hence the name WLM. It lies some $D \sim 1$\,Mpc from both the Milky Way and Andromeda \citep{2011AJ....141..194G} and so is remarkably isolated. It has excellent HI data, photometry and stellar kinematics \citep[e.g.][]{2012ApJ...750...33L,2012AJ....143...47Z,2015AJ....149..180O}. Its HI distribution is smooth, apart from the presence of a small HI hole of size $\sim 0.46$\,kpc (\citealt{2007AJ....133.2242K}, and see Figure \ref{fig:real_rotcurve} second row, leftmost panel). Its hole has no measured expansion velocity and so, similarly to the HI hole in NGC 6822, is likely quite old. There is no evidence for significant non-circular motions in the gas and its rotation curve is quite smooth \citep[e.g.][]{2015AJ....149..180O}. WLM appears to be relatively quiescent at the present time, showing substantially lower than average star formation over the past 0.1\,Gyr: $\SFRN = 0.43 \pm 0.11$ \citep{2012AJ....143...47Z}.

\vspace{2mm}
\noindent
{\bf IC 1613: }
IC 1613 was first discovered by \citet{1929AN....234..407B}. It is a near face-on dwarf irregular on the edge of the Local Group, some $\sim 740$\,kpc away \citep{2013ApJ...773..106S}. It has been mass modelled numerous times in the literature before \citep[e.g.][]{1989AJ.....98.1274L,2015AJ....149..180O}, while many authors have noted the clumpy nature of its ISM, with substantial HI bubbles and shells \citep{2002A&AT...21..223L,2006A&A...448..123S}. The most prominent of these has a size of $\sim 1$\,kpc and a large expansion velocity of $\sim 25$\,km/s. Its rotation curve also has a strange morphology, with two notable dips (see Figure \ref{fig:real_rotcurve}, second row, second panel). We discuss this further in \S\ref{sec:results}. IC 1613 formed stars over the past 0.1\,Gyr at a rate close to the mean over a Hubble time: $\SFRN = 0.81 \pm 0.25$ \citep{2012AJ....143...47Z}. This is substantially higher than WLM, DDO 101 and NGC 6822 and consistent with the violent appearance of its HI column density map.

\vspace{2mm}
\noindent
{\bf DDO 101: }
DDO 101 (also called UGC 6900) is substantially more distant than the other galaxies. Its distance is typically assumed to be $D_{\rm DDO101} = 6.4$\,Mpc \citep[e.g.][]{2015AJ....149..180O}, however the uncertainties on $D_{\rm DDO101}$ are very large. It it too far away for an accurate tip-of-the-red-giant-branch or Cepheid distance measurement, and so its distance must be determined instead from matching its peak rotation velocity to the Tully-Fisher relation \citep{1977A&A....54..661T}. This introduces large uncertainties both because of the intrinsic scatter in the Tully-Fisher relation, particularly at low peak rotation velocity \citep[e.g][]{2016MNRAS.455.3136M}, but also because its measured peak rotation velocity is only a lower bound on the true maximum. According to the NASA/IPAC Extragalactic Database (NED), its distance could lie in the range $6 < D_{\rm DDO101}/{\rm Mpc} < 16$, while its cosmological `Hubble flow' distance is 12.9\,Mpc (assuming the latest cosmological parameters from \citealt{2013arXiv1303.5076P}). The earliest literature on DDO 101 that we could find are its listing in the \citet{1973ugcg.book.....N} and \citet{1988cng..book.....T} catalogues. It has had its neutral hydrogen mapped by \citet{1990ApJS...72..245S} and most recently by Little THINGS \citep{2015AJ....149..180O}. Its star formation rate over the past 0.1\,Gyr is substantially lower than its average over a Hubble time, indicating that it is not currently starbursting (\citealt{2012AJ....143...47Z} estimate $\SFRN = 0.08 \pm 0.02$). DDO 101 is one of the few galaxies in the Little THINGS survey that has a steeply rising rotation curve and this is our motivation for including it here. \citet{2015AJ....149..180O} note that as a result, it is one of the few galaxies that appears to be well fit by an NFW profile.

\begin{table*}
\resizebox{\textwidth}{!}{
\begin{tabular}{L{1.4cm} | c c c c c c c | c c c c | c}
\hline
\hline
{\bf Galaxy} \vspace{1mm} & ${\mathbf v_{\rm max}}$ & ${\mathbf i}$ & $\mathbf{D}$ & $\mathbf{M_*}$ & $\mathbf{M_{\rm gas}}$ & $\mathbf{R_*}$ & $\mathbf{R_{\rm gas}}$ & $\mathbf{[R_{\rm min}, R_{\rm max}]}$ & $\mathbf{M_{200}}$ & $\mathbf{c}$ & $\mathbf{\chi^2_{\rm red}}$ & {\bf Refs.} \\
& (km/s) & ($^\circ$) & (kpc) & $({\rm M}_\odot)$ & $({\rm M}_\odot)$ & (kpc) &  (kpc) & (kpc) & $({\rm M}_\odot)$ &  &  & \\
\hline
\hline
\rowcolor{gray!40}NGC 6822 & 55 & 65 (inner); 75 (outer) & $490\pm 40$ & $7.63 \pm 1.9 \times 10^7$ & $17.4\times 10^7$ & 0.68 & 1.94 & $[2.5,-]$ & $2^{+0.2}_{-0.3} \times 10^{10}$ & $15.1^{+1.8}_{-0.8}$ & 0.37 & 1,2,3 \\ [3ex]

WLM & 39 & $74 \pm 2.3$ & $985\pm 33$ & $1.62 \pm 0.4 \times 10^7$ & $7.9 \times 10^7$ & $0.75$ & $1.04$ & $[0,-]$ & $8.3_{-2.2}^{+2.1}\times 10^9$& $17^{+3.9}_{-2.2}$ & 0.27 & 3,4,5,6 \\ [3ex]

\rowcolor{gray!40}IC 1613 & 20 & $39.4 \pm 2.29$ & $740 \pm 10$ & $1.5 \pm 0.5 \times 10^7$ & $8 \times 10^7$ & 0.65 & 1.29 & $[1.9,-]$ &  $4.7^{+1.2}_{-0.98} \times 10^8$ & $21.8_{-5.4}^{+5.3}$ & 0.13 & 3,6 \\ [1ex]
\rowcolor{gray!40} & 41 & $15$ & & & & & & $[1.9,-]$ & $7.75_{-2}^{+4}\times 10^9$ & $21.7_{-5.5}^{+5.5}$ & 0.32 & \\ [3ex]

DDO 101 & 65 & $52.4 \pm 1.7$ & 6,400 & $6.54 \pm 1 \times 10^7$ & $3.48 \times 10^7$ & 0.58 & 1.01 & $[0,-]$ & $5.2^{+0.6}_{-0.4} \times 10^{10}$ & $28.9_{-1.3}^{+0.6}$ & 7.2 & 3,6 \\ [1ex]
 & & & 12,900 & $26.6 \pm 4 \times 10^7$ & $14.13 \times 10^7$ & 1.16 & 2.03 & $[0,-]$ & $3.0^{+0.4}_{-0.2} \times 10^{10}$ & $28.3_{-2.2}^{+1.1}$ & 1.92 & \\ [1ex]
 & & & 16,000 & $40.9 \pm 6 \times 10^7$ & $21.75 \times 10^7$ & 1.45 & 2.5 & $[0,-]$ & $2.7^{+0.5}_{-0.2} \times 10^{10}$ & $27.6_{-3.3}^{+1.6}$ & 1.1 & \\ [1ex]
\hline
\hline
\end{tabular}
}
\caption{Four isolated dwarf irregular galaxies with excellent literature data, chosen to span a range of rotation curve shapes. The first column gives the galaxy name. Columns 2-7 give the data for that galaxy: the peak asymmetric drift corrected rotation curve velocity $v_{\rm max}$; the inclination angle $i$ in degrees (with formal 1-$\sigma$ errors; see \S\ref{sec:datacubes} for a discussion of the validity of these); the distance to the galaxy $D$; the stellar mass, with errors, $M_*$; the total gas mass $M_{\rm gas}$; and the exponential stellar and gas disc scale lengths $R_*$ and $R_{\rm gas}$, respectively. Column 8 gives the radial range used in the fit to the rotation curve $[R_{\rm min}, R_{\rm max}]$ (`$-$' indicates that $R_{\rm max}$ is set to the outermost data point). Columns 9-10 give the marginalised dark matter halo parameters: the virial mass $M_{200}$ and concentration parameter $c$, with 68\% confidence intervals. Column 11 gives the reduced $\chi^2_{\rm red}$ of the fit (we discuss how to interpret these $\chi^2_{\rm red}$ values in \S\ref{sec:resultsdata}). Finally, column 12 gives the data references for that galaxy as follows: 1: \citet{1884AN....110..125B}; 2: \citet{2003MNRAS.340...12W}; 3: \citet{2012AJ....143...47Z}; 4: \citet{2011AJ....141..194G}; 5: \citet{2012ApJ...750...33L}; 6: \citet{2015AJ....149..180O}. For IC 1613, there are two entries corresponding to different inclination angles $i$ (see \S\ref{sec:IC1613}), while for DDO 101 there are three corresponding to different distances $D$ (see \S\ref{sec:DDO101}).}
\label{tab:data}
\end{table*}

\subsection{Extracting rotation curves from HI data cubes}\label{sec:barolo}

We derived the rotation curves from the HI datacubes using the publicly available software \Barolo\footnote{{\tt http://editeodoro.github.io/Bbarolo/}.} \citep{2015MNRAS.451.3021D}. \Barolo\ fits tilted-ring models directly to the datacube by building artificial 3D data and minimising the residuals, without explicitly extracting velocity fields. This ensures full control of the observational effects and in particular a proper account of beam smearing that can strongly affect the derivation of the rotation velocities in the inner regions of dwarf galaxies \citep[see e.g.][]{1997ApJ...491..140S}. \Barolo\ fits up to 9 parameters for each ring in which the galaxy is decomposed, namely: central coordinates; systemic velocity; inclination ($i$); position angle (p.a.); HI density; HI thickness; rotation velocity ($v_{\rm rot}$); and velocity dispersion ($\sigma_{\rm HI}$). 

To derive our curves, we made the following assumptions. We fixed the centre of all rings to the centre of the galaxies obtained from the literature (\citealt{2012AJ....144..134H} for the Little THINGS galaxies and \citealt{2012AJ....144....4M} for NGC 6822), and we fixed the systemic velocity to the value calculated as:

\begin{equation}
V_{\rm sys}=0.5\left({\rm V}20_{\rm app}+{\rm V}20_{\rm rec}\right)
\end{equation}
where V20 is the velocity where the flux of the global HI profile reaches the 20\% with respect to the flux peak, while `app' and `rec' indicate the approaching and receding halves of the galaxy. We did not fit the HI density, but instead normalised the flux locally to the value of the total HI map. For a full description of this normalisation technique, see \citet{2015MNRAS.451.3021D}. In all cases, we considered the disk thin and fixed the scale height to 100\,pc, constant in radius. This assumption will be improved in a forthcoming work where we will take into account the flaring of the HI disc using a self-consistent method based on vertical hydrostatic equilibrium (Iorio et al., in preparation). Given the above assumptions, we are left with four fitting parameters: $i$, p.a., $v_{\rm rot}$ and $\sigma_{\rm HI}$.

In order to obtain a good fit of the kinematics, it is important to start with reasonable initial guesses for the inclination $i$ and the position angle p.a. We use \Barolo\ to estimate these initial guesses by fitting the total HI map. We then estimated rotation and dispersion in two stages. First \Barolo\ makes a fit leaving the four parameters free.
Then it fixes the geometrical parameters, regularising them with a polynomial and performing a new fit of $v_{\rm rot}(R)$ and $\sigma_{\rm HI}(R)$ alone.

The HI data for WLM, IC\,1613 and DDO\,101 were obtained from the publicly available archive of the Little THINGS survey\footnote{{\tt https://science.nrao.edu/science/surveys/Little THINGS}}, while the HI datacube of NGC\,6822 was kindly provided to us by Erwin de Blok. For WLM and IC\,1613, we used natural weighted data smoothed to a resolution of 25 arcseconds that represents a good compromise between the number of resolution elements and the enhancement of the galaxy signal. The extent of DDO\,101 on the sky is very small so we used the robust weighted data at the original resolution of about 8 arcsecond without further smoothing. Finally, for NGC 6822 we smoothed the original cube from a resolution of 42x12 to 43x30 arcsec.

Except for the case of NGC 6822, the data did not show any clear radial trends in the geometrical parameters, so we fitted them with a constant value; we report the best-fit values of $i$ with formal error bars in Table \ref{tab:data}. For NGC 6822, we found that the inclination rises from about 65 degrees in the centre to 70 degrees in outskirts.

The final rotation curves were corrected for asymmetric drift (\S\ref{sec:asymcorrect}), fitting the $\Sigma_{\rm g}\sigma^2_{\rm g}$ data with a functional form (see Appendix \ref{app:asymcorrect} for further details and tests of our asymmetric drift correction).

\section{The rotation curve fitting method}\label{sec:rotmethod}

\subsection{The mass model}\label{sec:massmodel}

We decompose the circular speed curve into contributions from stars, gas and dark matter: 

\begin{equation}
v_c^2 = v_*^2 + v_{\rm gas}^2 + v_{\rm dm}^2
\end{equation}
where $v_*$ and $v_{\rm gas}$ are the contributions from stars and gas, respectively, and $v_{\rm dm}$ is the dark matter contribution. We assume that both the stars and gas are well-represented by exponential discs: 

\begin{equation} 
v_{*/{\rm gas}}^2 = \frac{2 G M_{*/{\rm gas}}}{R_{*/{\rm gas}}} y^2 \left[I_0(y) K_0(y) - I_1(y) K_1(y)\right]
\label{eqn:vcstargas}
\end{equation}
where $M_{*/{\rm gas}}$ is the mass of the star/gas disc, respectively; $R_{*/{\rm gas}}$ is the exponential scale length; $y = R/R_{*/{\rm gas}}$ is a dimensionless radius parameter; and $I_0, I_1, K_0$ and $K_1$ are Bessel functions \citep{1987gady.book.....B}. For the mock simulation data, we input the best-fit values of $M_*, M_{\rm gas}, R_*$ and $R_{\rm gas}$. For the real data, we use the measured values of $M_*$ (from stellar population synthesis modelling) and $M_{\rm gas}$. We use either reported single exponential fits to the surface density profile (for NGC 6822); or we fit a single exponential to the stellar and gas surface densities reported in \citet{2012AJ....143...47Z} and \citet{2015AJ....149..180O}, respectively. We fix the values of $R_*$ and $R_{\rm gas}$ in advance of running our Markov Chain Monte Carlo (MCMC) models (see \S\ref{sec:mcmc}). All values used are reported in Table \ref{tab:data}. To check the validity of equation \ref{eqn:vcstargas} for the gas, we ran tests where we calculated $v_{\rm gas}$ directly from the gas surface density data (i.e. not assuming an exponential) for both WLM and IC 1613; the differences as compared to using the exponential fit were negligible.

For the dark matter profile, we use the \coreNFW\ profile from R16: 

\begin{equation}
M_{\rm cNFW}(<r) = M_{\rm NFW}(<r) f^n
\label{eqn:McNFW}
\end{equation}
where $M_{\rm NFW}(<r)$ is the usual NFW enclosed mass profile \citep{1996ApJ...462..563N}:

\begin{equation} 
M_{\rm NFW}(<r) = M_{200} g_c \left[\ln\left(1+\frac{r}{r_s}\right) - \frac{r}{r_s}\left(1 + \frac{r}{r_s}\right)^{-1}\right]
\label{eqn:MNFW}
\end{equation}
where $M_{200}$; $c$; $r_s$; $g_c$; $\rho_{\rm crit} = 128.2$\,M$_\odot$\,kpc$^{-3}$; and $\Delta = 200$ are as in equation \ref{eqn:rhoNFW}.

The function $f^n$ generates a shallower profile below a `core radius' $r_c$: 

\begin{equation} 
f^n = \left[\tanh\left(\frac{r}{r_c}\right)\right]^n
\end{equation}
where the parameter $0 < n \le 1$ controls how shallow the core becomes ($n=0$ corresponds to no core; $n=1$ to complete core formation). The parameter $n$ is tied to the total star formation time\footnote{More precisely, the total {\it duration} of star formation, not to be confused with the star formation depletion timescale $t_{\rm dep}=\Sigma_{\rm gas}/\Sigma_{\rm SFR}$ \citep[e.g.][]{Bigiel2011}.} $t_{\rm SF}$:
\begin{equation} 
n = \tanh(q) \,\,\,\, ; \,\,\,\, q = \kappa \frac{t_{\rm SF}}{t_{\rm dyn}}
\end{equation} 
where $t_{\rm dyn}$ is the circular orbit time at the NFW profile scale radius $r_s$:
\begin{equation} 
t_{\rm dyn} = 2\pi \sqrt{\frac{r_s^3}{G M_{\rm NFW}(r_s)}}
\end{equation}
and $\kappa = 0.04$ is a fitting parameter (see R16). For the isolated dwarfs that we consider here, we assume $t_{\rm SF} = 14$\,Gyrs such that they have formed stars continuously for a Hubble time.

The `core size'\footnote{ 
Note that the true `size' of the dark matter core is somewhat arbitrary and depends on what definition we use \citep[see e.g. the discussion in][]{Goerdt:2006rw}. From R16, their figure 4, the onset of the dark matter core occurs visually at $\sim R_{1/2}$ and hence we refer throughout this paper to the dark matter core being of `size' $\sim R_{1/2}$. To reproduce this behaviour with the \coreNFW\ model, however, we require that our `core size' parameter $r_c$ is nearly twice $R_{1/2}$, as in equation \ref{eqn:etarc}.} is set by the projected half stellar mass radius of the stars $R_{1/2}$:

\begin{equation} 
r_c = \eta R_{1/2}
\label{eqn:etarc}
\end{equation} 
For an exponential disc, $R_{1/2} = 1.68 R_*$. By default, we set $\eta = 1.75$ since this gives the best match to the simulations in R16. However, as discussed in R16, there could be some scatter in $\eta$ due to varying halo spin, concentration parameter and/or halo assembly history. For this reason, in Appendix \ref{app:etafree} we explore allowing $\eta$ to vary freely over the range $0 < \eta < 5$ when fitting data for WLM. There, we show that this further limits our ability to measure the halo concentration parameter $c$ and slightly inflates our errors on $M_{200}$, as might be expected, but is otherwise benign. Interestingly, we find $\eta_{\rm WLM} = 2.4_{-0.52}^{+0.78}$ at 68\% confidence, consistent with our favoured $\eta = 1.75$.

Finally, note that the ratio of a completely cored \coreNFW\ rotation curve (with $n=1$) to an NFW rotation curve is given by:

\begin{equation} 
\frac{v_{{\rm dm},{\rm cNFW}}}{v_{{\rm dm},{\rm NFW}}} = \sqrt{\tanh\left({\frac{R}{1.75 R_{1/2}}}\right)}
\label{eqn:coreNFWrotratio}
\end{equation}
where we now write $v_{\rm dm}$ as a function of cylindrical coordinate $R$ in the disc plane.

From equation \ref{eqn:coreNFWrotratio}, we see that even at $R = 2 R_{1/2}$, the \coreNFW\ rotation curve has $\sim 90$\% of the amplitude of the equivalent NFW rotation curve. Since for our simulations, the dark matter dominates the total enclosed mass at all radii, the total rotation curve $v_c(R) \simeq v_{\rm dm}(R)$. Thus, the dark matter cores in R16, although visually of a size $\sim R_{1/2}$, will affect the full rotation curve out to $\sim 2 R_{1/2}$.

\subsection{Fitting the mass model to data \& our choice of priors}\label{sec:mcmc}

We fit the above mass model to the data using the \EMCEE\ affine invariant Markov Chain Monte Carlo (MCMC) sampler from \citet{2013PASP..125..306F}. We assume uncorrelated Gaussian errors such that the Likelihood function is given by $\mathcal{L} = \exp(-\chi^2/2)$. We use 100 walkers, each generating 1500 models and we throw out the first half of these as a conservative `burn in' criteria. We explicitly checked that our results are converged by running more models and examining walker convergence. All parameters were held fixed except for the dark matter virial mass $M_{200}$; the concentration parameter $c$; and the total stellar mass $M_*$. We assume a flat logarithmic prior on $M_{200}$ of $8 <  \log_{10}\left[M_{200}/{\rm M}_\odot\right] < 11$; a flat linear prior on $c$ of $14 < c < 30$ and a flat linear prior on $M_*$ over the range given by stellar population synthesis modelling, as reported in Table \ref{tab:data}. For the mock simulation data and the real data, we assume an error on $M_*$ of $25\%$ unless a larger error than this is reported in the literature \citep[e.g.][]{2012AJ....143...47Z,2015AJ....149..180O}. The generous prior range on $c$ is set by the cosmic mean redshift $z=0$ expectation value of $c$ at the extremities of the prior on $M_{200}$ \citep{2007MNRAS.378...55M}; we explore our sensitivity to this choice of prior in Appendix \ref{app:cprior}. For each galaxy, we fit data over a range $[R_{\rm min}, R_{\rm max}]$ as reported in Table \ref{tab:data}. Where we write `--' for $R_{\rm max}$, this means that $R_{\rm max}$ is set by the outer edge of the rotation curve data.

\section{Results}\label{sec:results} 

\begin{figure}
\begin{center}
\includegraphics[width=0.49\textwidth]{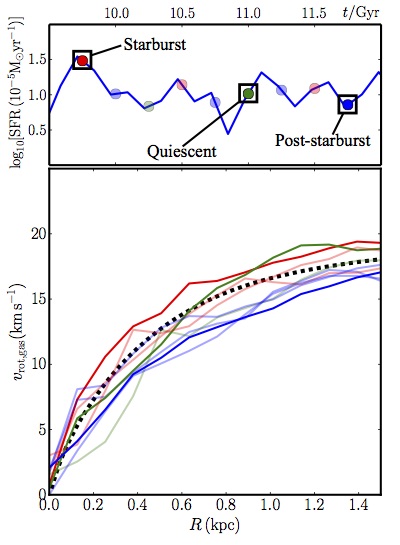}
\caption{Asymmetric drift corrected rotation curves (c.f. \S\ref{sec:asymcorrect}) for our $M_{200} = 10^9$\,M$_\odot$ simulation at 9 regularly spaced time intervals over the starburst cycle (bottom panel). The upper panel shows the star formation rate as a function of time; the lower panel shows the gas rotation curves at the times marked by the circles in the upper panel. The red circles correspond to a `starburst' phase in the cycle; the blue to `post-starburst'; and the green to a more `quiescent' phase. In \S\ref{sec:mockfit}, we will fit models to the three times marked by the black squares. Notice that there is a general trend that when the star formation rate is increasing, the rotation curve amplitude is higher (red), while when it is decreasing, the rotation curve amplitude is lower (blue). The black dashed line marks the true rotation curve as calculated from the gravitational potential.}
\label{fig:mock_common} 
\end{center}
\end{figure}

\subsection{Fitting models to ideal mock data}\label{sec:mock}

In this section, we first consider mock data that are `as good as it gets'; that is, we assume that we can perfectly inclination and asymmetric drift correct the rotation curves, as described in \S\ref{sec:simulations}. We then explore how well we can recover the dark matter halo mass $M_{200}$ and concentration parameter $c$ when fitting the mass model described in \S\ref{sec:rotmethod} to these mock data. (We will explore how well we can inclination and asymmetric drift correct mock HI data cubes in \ref{sec:datacubes}.)

\subsubsection{Starburst-induced variance in the rotation curve}\label{sec:mockvar}

Before fitting the mock rotation curves, let us first take a look at the time evolution of the mock galaxy rotation curve through the star burst cycle. In Figure \ref{fig:mock_common}, we show asymmetric drift corrected rotation curves (c.f. \S\ref{sec:asymcorrect}) for our $M_{200} = 10^9$\,M$_\odot$ simulation at 9 regularly spaced time intervals over the starburst cycle (bottom panel). The upper panel shows the star formation rate as a function of time; the lower panel shows the gas rotation curves at the times marked by the circles in the upper panel. The red circles correspond to a `starburst' phase in the cycle; the blue to `post-starburst'; and the green to a more `quiescent' phase. In \S\ref{sec:mockfit}, we will fit models to the three times marked by the black squares. 

Notice that there is a general trend that when the star formation rate is increasing, the rotation curve amplitude is higher (red), while when it is decreasing, the rotation curve amplitude is lower (blue). At quiescence, the rotation curve lies in between these extremes (green), in good agreement with the true rotation curve\footnote{Since the stars and gas are sub-dominant to the dark matter at all radii, for this `true rotation curve', we assume spherical symmetry such that $v_c^2 = GM_{\rm tot} / R$, where $M_{\rm tot}$ is the total enclosed mass, as calculated from the gravitational potential.} (black dashed line). Such a movement in the asymmetric drift corrected rotation curve occurs continuously throughout the starburst cycle. If real galaxies behave similarly to this, then we expect them to be equally often in all three phases, though the most extreme departures from quiescence will be more rare. We now consider how such a variation in the rotation curve affects rotation curve modelling.

\subsubsection{Fitting mock rotation curves through the starburst cycle}\label{sec:mockfit}

\begin{figure*}
\begin{center}
\includegraphics[width=\textwidth]{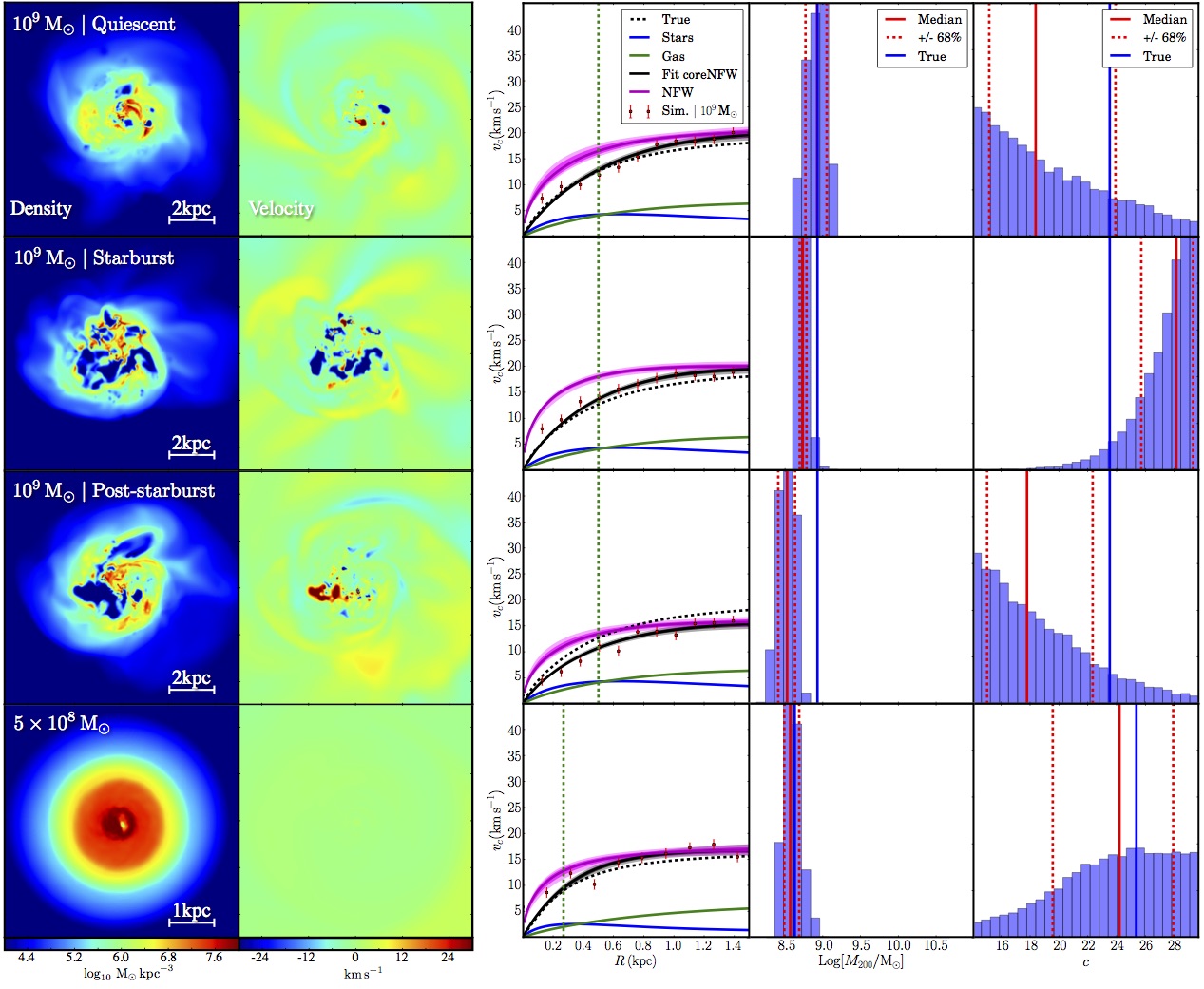}
\caption{Fitting the model described in \S\ref{sec:rotmethod} to ideal mock rotation curve data. From left to right, the columns show the gas density viewed face-on; the mean vertical velocity of the gas for this same face-on view; the fitted asymmetric drift corrected rotation curve; and the resultant constraints on the dark matter virial mass $M_{200}$ and concentration parameter $c$. On the rotation curve plots, we mark the projected half stellar mass radius $R_{1/2}$ by the vertical green dashed line; the mock rotation curve data with errors (red data points); the true model rotation curve (black dashed line); the star (blue) and gas (green) contribution to the rotation curve; and the median (black); 68\% (dark grey) and 95\% (light grey) confidence intervals of our fitted model rotation curves. The magenta line and dark/light magenta bands show what the median, 68\% and 95\% confidence intervals of our model rotation curves would look like if we switched off cusp-core transformations (i.e. if we apply the NFW profiles that correspond to our fitted \coreNFW\ profiles). The fourth and fifth panels from the left show histograms of our recovered $M_{200}$ and $c$; the true answers are marked by the vertical blue lines.}
\label{fig:mock_rotcurve} 
\end{center}
\end{figure*}

In Figure \ref{fig:mock_rotcurve}, we fit the model described in \S\ref{sec:rotmethod} to our mock rotation curve data. We analyse the $10^9$\,M$_\odot$ simulation at three times, marked as `quiescent'; `starburst' and `post-starburst' on Figure \ref{fig:mock_common} (top three rows). We also analyse the $5 \times 10^8$\,M$_\odot$ simulation at a simulation time of 14\,Gyrs (bottom row). From left to right, the columns show the gas density viewed face-on; the mean vertical velocity of the gas for this same face-on view; the fitted asymmetric drift corrected rotation curve; and the resultant constraints on the dark matter virial mass $M_{200}$ and concentration parameter $c$. On the rotation curve plots, we mark the projected half stellar mass radius $R_{1/2}$ by the vertical green dashed line; the mock rotation curve data with errors (red data points); the true model rotation curve (black dashed line); the star (blue) and gas (green) contribution to the rotation curve; and the median (black); 68\% (dark grey) and 95\% (light grey) confidence intervals of our fitted model rotation curves. The magenta line and dark/light magenta bands show what the median, 68\% and 95\% confidence intervals of our model rotation curves would look like if we switched off cusp-core transformations (i.e. if we apply the NFW profiles that correspond to our fitted \coreNFW\ profiles). The fourth and fifth panels from the left show histograms of our recovered $M_{200}$ and $c$; the true answers are marked by the vertical blue lines.

Firstly, notice that in all simulations there are HI bubbles being blown through the disc by stellar feedback (Figure \ref{fig:mock_rotcurve}, leftmost panels). These typically reach a velocity $\sim 30$\,km/s (Figure \ref{fig:mock_rotcurve}, second column) in excellent agreement with data for real isolated dwarfs (see Table \ref{tab:data} and \S\ref{sec:resultsdata}). The `quiescent' mock (top row) shows very little activity, resulting in a rotation curve that is closely matched to the underlying circular speed curve $v_{\rm rot,gas} \sim v_c$. Our fit returns a minimum reduced $\chi^2_{\rm red} = 1.79$, corresponding to a good representation of the data. We obtain an excellent recovery of $M_{200}$ within our quoted uncertainties (compare the vertical blue line with the histogram in the fourth panel from left, upper row). However, it is substantially harder to recover the halo concentration $c$ and the constraints are much poorer (see rightmost panel, upper row).

The second row of Figure \ref{fig:mock_rotcurve} shows our results for the `starburst' mock. At this output time, the gas is in a highly turbulent state, and features many fast-moving HI bubbles of size $\sim 0.5 - 1$\,kpc and expansion velocity $\sim 30$\,km/s. This causes a substantially steeper rise in the inner rotation curve. Interestingly, our \EMCEE\ fit skirts between the inner and outer rotation curve data points, leading to a highly biased concentration parameter $c$ that pushes on our prior, and an {\it underestimate} of $M_{200}$. The minimum reduced $\chi^2_{\rm red} = 2.4$ is noticeably poorer than for the quiescent case. This is because the rotation curve within $R_{1/2}$ (vertical green dashed line) rises more steeply than our model ensemble. If we restrict our fit to $R < R_{1/2}$, then we find that we systematically overestimate $M_{200}$, as might naively be expected for a rotation curve that rises too steeply.  

The third row of Figure \ref{fig:mock_rotcurve} shows our results for the `post-starburst' mock. Like the starburst mock, it is similarly out of equilibrium, with substantial HI bubbles. However, instead of many smaller HI bubbles, these have agglomerated into one enormous outflow. The rotation curve is now systematically shallower than in quiescence, even out to the outermost rotation curve data point at $R = 1.5$\,kpc. When fitting our mass model to these mock data, this leads to an underestimate of $M_{200}$ by $\sim$ half a dex. However, the minimum reduced $\chi^2_{\rm red} = 1.0$ indicates an excellent fit. This demonstrates that we cannot rely on $\chi^2_{\rm red}$ alone as an indicator that the rotation curve is out of equilibrium. Instead, we must look for evidence of fast-moving and substantial HI bubbles in the HI velocity field, with associated star formation activity over the past $\sim 100$\,Myrs (see Figure \ref{fig:mock_common}). 

\subsubsection{The importance of cusp-core transformations}\label{sec:nfwfit}

In the third column of Figure \ref{fig:mock_rotcurve}, we illustrate the importance of properly accounting for cusp-core transformations driven by stellar feedback. The magenta bands show the rotation curves that would arise if we `undid' the cusp-core transforms (i.e. if we insert the best fitting \coreNFW\ profile parameters into the NFW profile and calculate the resulting rotation curves). Notice that these all rise substantially more steeply than the data, as expected. If we fit NFW profiles to these mock data instead of \coreNFW, we become biased towards extremely low concentration parameters inconsistent with cosmological expectations. The mass $M_{200}$ is, however, still correctly recovered so long as the rotation curve data extend far enough out. If the rotation curve does not extend to the point where it becomes flat, then when fitting NFW instead of \coreNFW, we become biased also towards low $M_{200}$ (see Appendix \ref{app:nfwfit}). 

\subsubsection{Fitting rotation curves at the edge of galaxy formation}

In the bottom row of Figure \ref{fig:mock_rotcurve}, we fit the rotation curve for our lower mass mock with $M_{200} = 5 \times 10^8$\,M$_\odot$. Despite having just about the lowest mass possible for a galaxy that can continue to form stars for a Hubble time (see discussion in R16), our mock displays a rather smooth rotation curve that -- once corrected for asymmetric drift -- corresponds well with the underlying potential. As a result, we obtain an excellent recovery of $M_{200}$. This suggests that we should be able to recover $M_{200}$ and $c$ from even very tiny dwarf irregulars like LeoT or Aquarius \citep[e.g.][]{2003ApJ...592..111Y,2007ApJ...656L..13I,2008MNRAS.384..535R}.

\subsection{The effect of the starburst cycle on the stellar kinematics}\label{sec:stellarkin}

\begin{figure}
\begin{center}
\hspace{-5mm}
\includegraphics[width=0.49\textwidth]{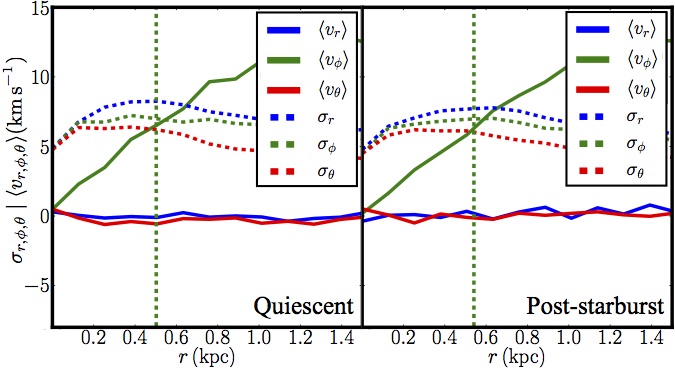}
\caption{The radial (blue) and angular (green, $\phi$; red, $\theta$) components of the stellar velocity dispersion in spherical polar coordinates ($\sigma_{r,\phi,\theta}$; dashed lines) and the mean streaming velocity in these coordinates ($\langle v_{r,\phi,\theta}\rangle$; solid lines), as indicated in the legend, for our $M_{200} = 10^9$\,M$_\odot$ mock in quiescence (left) and post-starburst (right). The vertical green dashed line marks the projected half stellar mass radius $R_{1/2}$ in both cases.}
\label{fig:mock_stars} 
\end{center}
\end{figure}

There are now a subset of isolated dwarfs with stellar kinematic data \citep[e.g.][]{2014ApJ...789...63A,2014MNRAS.439.1015K}. For this reason, it is interesting to ask whether the stars are also pushed out of equilibrium by stellar feedback. In Figure \ref{fig:mock_stars}, we show the radial (blue) and angular (green, $\phi$; red, $\theta$) components of the stellar velocity dispersion in spherical polar coordinates ($\sigma_{r,\phi,\theta}$; dashed lines) and the mean streaming velocity in these coordinates ($\langle v_{r,\phi,\theta}\rangle$; solid lines), as indicated in the legend, for our $M_{200} = 10^9$\,M$_\odot$ mock in quiescence (left) and post-starburst (right). Notice that the two plots are very similar. In both cases, dispersion dominates over rotation inside $R_{1/2}$ (vertical green line), while the opposite is true beyond $R_{1/2}$. However, while the two cases are remarkably similar, the post-starburst system has a slightly larger projected half stellar mass radius while its dispersions rise slightly more slowly and with lower radial anisotropy. Using an analysis similar to that presented in R16, we have verified that these small changes are not sufficient to substantially bias Jeans modelling of the stars by more than $\sim 1$\,km/s (assuming perfect data). 

We may be tempted to conclude from the above analysis that stars are a more robust probe of the underlying potential than the gas. However, as emphasised in R16, the stars give a reliable estimate of the mass only within $\sim R_{1/2}$ \citep[e.g.][]{2009ApJ...704.1274W,2010MNRAS.406.1220W,2016arXiv160304443C}. Yet this is precisely where stellar feedback affects the gravitational potential (R16; and see \citealt{2015arXiv150202036O}), making it challenging to obtain a robust estimate of $M_{200}$. By contrast, HI gas traces the gravitational potential much further out, often to the point where the rotation curve becomes flat. Thus, stellar kinematic and HI data remain complementary. We defer a full analysis of the efficacy of combined stellar and HI kinematics to future work.

\subsection{Deprojecting mock HI datacubes: how well can we recover the inclination?}\label{sec:datacubes}

\begin{figure*}
\begin{center}
\includegraphics[width=0.99\textwidth]{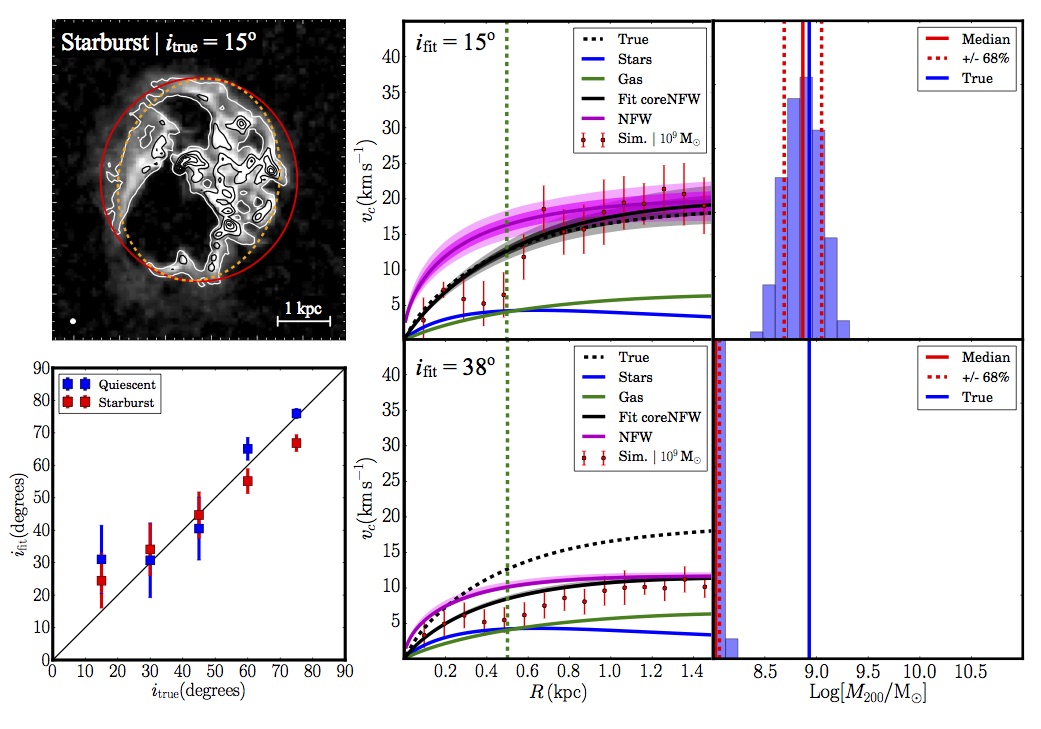}
\caption{Recovering the inclination and rotation curve from mock HI datacubes. The top left panel shows the surface density of our `starburst' mock dwarf viewed at an inclination of $i_{\rm true} = 15^\circ$. The red circle marks an ellipse showing this correct inclination ($i_{\rm fit} =  i_{\rm true} = 15^\circ$); the orange dashed circle shows instead $i_{\rm fit} = 38^\circ$. The bottom left panel plots the true inclination $i_{\rm true}$ versus the recovered inclination $i_{\rm fit}$ from the \Barolo\ code (see \S\ref{sec:barolo}) for the `quiescent' mock (blue) and the `starburst' mock (red) at inclinations $i_{\rm true} = 15^\circ, 30^\circ, 45^\circ, 60^\circ, 75^\circ$. The error bars mark the 68\% confidence intervals generated from 100 different simulated observations of the same datacube (see text for details). The top right two panels show the recovered rotation curve from \Barolo\ for $i_{\rm fit} = i_{\rm true} = 15^\circ$, and the resulting histogram of $M_{200}$ from our model fits (the lines and symbols are as in Figure \ref{fig:mock_rotcurve}). The bottom right two panels show the same for $i_{\rm fit} = 38^\circ$, corresponding to a large systematic overestimate of the inclination angle. This leads to an artificially shallow rotation curve, with an amplitude that is suppressed by a factor $\sin(i_{\rm true})/\sin(i_{\rm fit}) \sim 0.4$.}
\label{fig:mock_inclination} 
\end{center}
\end{figure*}

In this section, we now consider how well we can recover the rotation curve from mock HI datacubes. The mock data are described in \S\ref{sec:mockcubes}, while extracting rotation curves from these using \Barolo\ is described in \S\ref{sec:barolo}. The results for the quiescent and starburst mocks are shown in Figure \ref{fig:mock_inclination}. The top left panel shows the surface density of the starburst mock viewed at an inclination of $i_{\rm true} = 15^\circ$. The red circle marks an ellipse showing this correct inclination ($i_{\rm fit} =  i_{\rm true} = 15^\circ$); the orange dashed circle shows instead $i_{\rm fit} = 38^\circ$. The bottom left panel plots the true inclination $i_{\rm true}$ versus the recovered inclination $i_{\rm fit}$ from the \Barolo\ code (see \S\ref{sec:barolo}) for the `quiescent' mock (blue) and the `starburst mock' (red) at inclinations $i_{\rm true} = 15^\circ, 30^\circ, 45^\circ, 60^\circ, 75^\circ$. The error bars mark the 68\% confidence intervals generated from 100 different simulated observations of the same datacube (recall that these are different because each realisation has different Gaussian random noise added to it; see \S\ref{sec:mockcubes}). The top right two panels show the recovered rotation curve from \Barolo\ for $i_{\rm fit} = i_{\rm true} = 15^\circ$, and the resulting histogram of $M_{200}$ from our model fits (the lines and symbols are as in Figure \ref{fig:mock_rotcurve}). The bottom right two panels show the same for $i_{\rm fit} = 38^\circ$, corresponding to a large systematic overestimate of the inclination angle.

Notice from the bottom left plot of $i_{\rm fit}$ versus $i_{\rm true}$ that it is more challenging to recover the correct inclination for the starburst mock as compared to the quiescent mock. This is not surprising since the former has significant HI bubbles that distort the HI column density map, making it more challenging to measure inclination. More striking, however, is the tendency for \Barolo\ to overestimate $i_{\rm fit}$ for $i_{\rm true} \simlt 30^\circ$. At $i_{\rm true} = 15^\circ$, \Barolo\ sometimes favours an inclination as high as $i_{\rm fit} \sim 40^\circ$. This will cause an artificial suppression of the rotation curve by a factor $\sin(i_{\rm true})/\sin(i_{\rm fit}) \sim 0.4$, as can be seen in the middle row of Figure \ref{fig:mock_inclination}. For $i_{\rm fit} = i_{\rm true} = 15^\circ$, the rotation curve is well-recovered (Figure \ref{fig:mock_inclination}, top right two panels). When fitting the \coreNFW\ profile to these rotation curve data, we obtain an excellent recovery of $M_{200}$ (Figure \ref{fig:mock_inclination}, top right panel). By contrast, however, for $i_{\rm fit} = 38^\circ$, the rotation curve rises extremely slowly (Figure \ref{fig:mock_inclination}, bottom right two panels). We are still able to obtain a fit using the \coreNFW\ profile, but $M_{200}$ now pushes on the lower bound of our prior, and is too low by over a dex. In \S\ref{sec:resultsdata}, we will present an example of a real dwarf irregular galaxy -- IC 1613 -- that appears to behave similarly to this low inclination starbursting mock dwarf.

The above suggests that to be certain that inclination errors are not substantial, we should avoid dwarf irregulars with $i_{\rm fit} \simlt 40^\circ$ and/or those with substantial fast-moving HI bubbles. A dwarf with a `best-fit' inclination of $i_{\rm fit} \sim 40^\circ$ could have a true inclination as low as $i_{\rm true} \sim 15^\circ$. 

Finally, notice that for $i_{\rm fit} = i_{\rm true} = 15^\circ$ the rotation curve recovered from \Barolo\ shows a statistically significant `dip' at $R \sim 0.4$\,kpc (top middle panel, Figure \ref{fig:mock_inclination}). In \S\ref{sec:resultsdata}, we shall see similar such features in NGC 6822 and IC 1613. In the region of the dip, the rotation velocity is poorly constrained due to a combination of the presence of a large HI hole, low inclination and the presence of strong non-circular motions. These effects are reduced with increasing inclination and the dip is gone for $i_{\rm true} = 60^\circ$. This explains why such dips are not seen in the ideally extracted mock rotation curves in Figure \ref{fig:mock_rotcurve}.

It is likely that using a combination of stellar surface density information and more sophisticated slicing on HI column density, \Barolo\ can be improved for low inclination and/or starbursting dwarfs. Indeed, we emphasise that the process we adopted here to estimate the inclinations is completely blind; an iterative procedure will likely yield improved results, particularly at high inclinations. We will consider this in future work where we will also consider the effect of radially varying disc thickness.

\subsection{Application to real data}\label{sec:resultsdata}

\begin{figure*}
\begin{center}
\includegraphics[width=\textwidth]{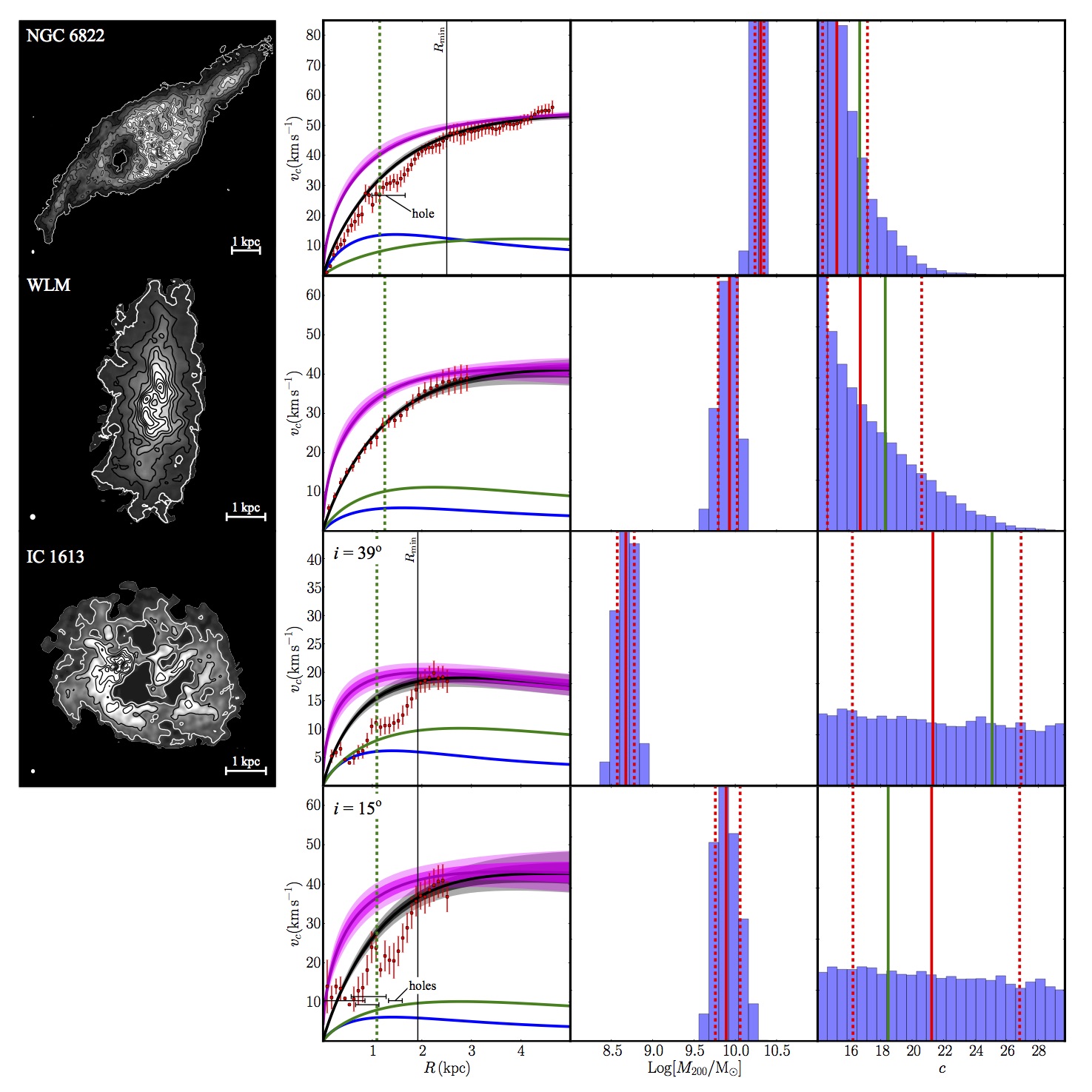}
\caption{Rotation curve models for three isolated dwarf irregular galaxies: NGC 6822 (top); WLM (middle); and IC 1613 (bottom). For IC 1613, we show results for two different inclination angles: $i = 39^\circ$ and $i = 15^\circ$, as marked. The columns show from left to right: the HI column density map with size-scale and beam-size marked; the rotation curve data; and histograms of $M_{200}$ and $c$ from our \EMCEE\ model chains. For the latter three panels, the lines and symbols are as in Figure \ref{fig:mock_rotcurve}. On the rotation curve plot (second column), we mark the position of HI holes seen in the column density map (leftmost column) of NGC 6822 and IC 1613. On the concentration parameter plot (rightmost column), we mark the cosmological mean $c_{\Lambda{\rm CDM}}$ (vertical green line), expected for a halo with the median $M_{200}$ from our model chains.}
\label{fig:real_rotcurve} 
\end{center}
\end{figure*}

\begin{figure*}
\begin{center}
\includegraphics[width=\textwidth]{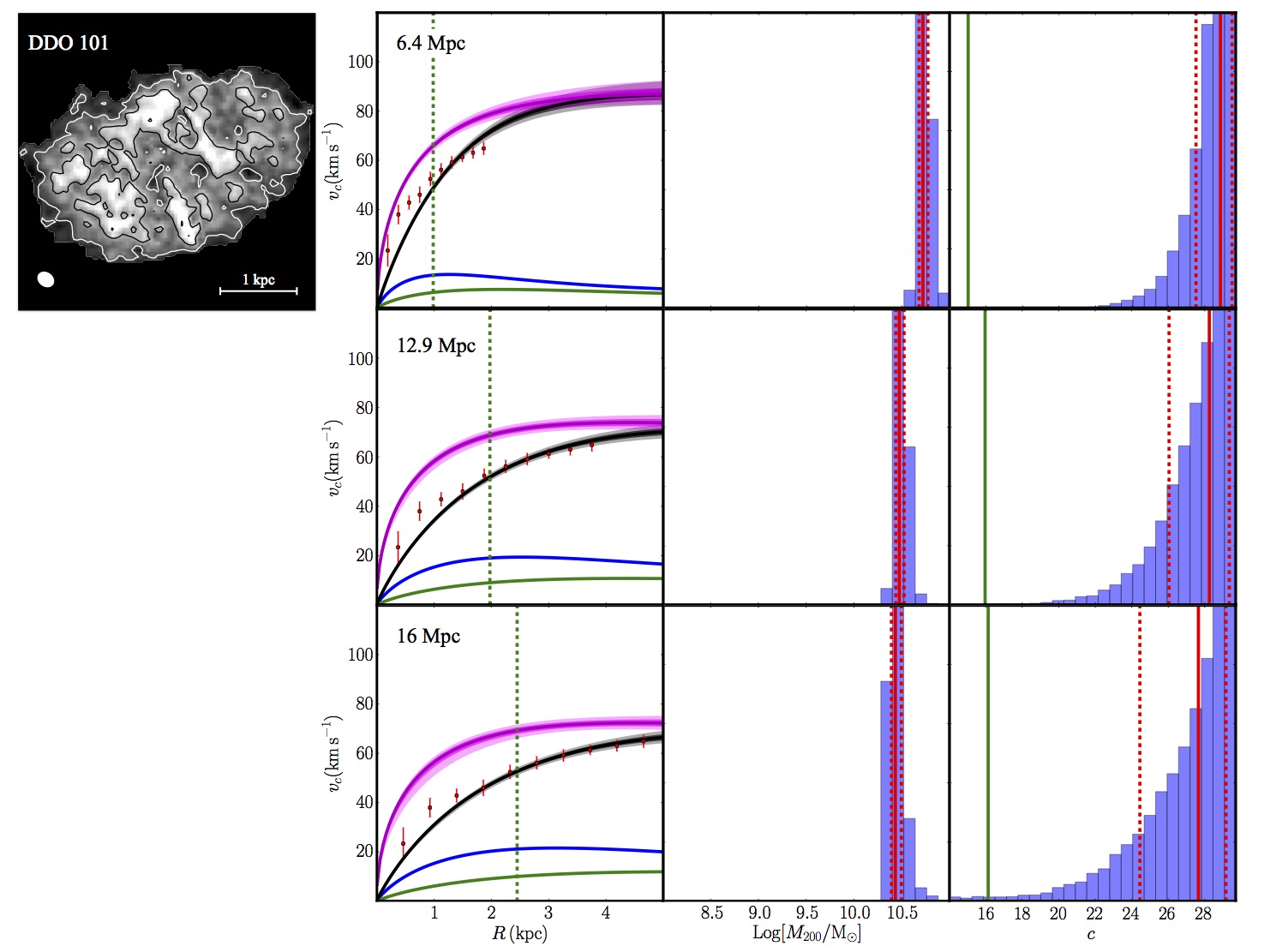}
\caption{Rotation curve models for DDO 101. From top to bottom, the rows show results assuming three different distances to the dwarf $D = 6.4$\,Mpc (as assumed previously in the literature); $D = 12.9$\,Mpc (the Hubble flow distance); and $D = 16$\,Mpc (the maximum distance reported in the literature). The columns lines and symbols are as in Figure \ref{fig:real_rotcurve}.}
\label{fig:real_rotcurve_ddo101} 
\end{center}
\end{figure*}

In this section, we apply our rotation curve fitting methodology (\S\ref{sec:rotmethod}) to real data for four isolated dwarf irregulars, chosen to span a range of interesting rotation curve morphologies: NGC 6822; WLM; IC 1613; and DDO 101. The data are described in detail in \S\ref{sec:data}, while our results are given in Figures \ref{fig:real_rotcurve} and \ref{fig:real_rotcurve_ddo101}.

\subsubsection{NGC 6822 \& WLM}

The top two rows of Figure \ref{fig:real_rotcurve} show our results for NGC 6822 and WLM. The columns show, from left to right: the HI column density map with size-scale marked; the rotation curve data; and histograms of $M_{200}$ and $c$ from our \EMCEE\ model chains. For the latter three panels, the lines and symbols are as in Figure \ref{fig:mock_rotcurve}. 

The rotation curve for NGC 6822 shows a clear dip just beyond the projected half stellar mass radius. This could owe to the prominent HI bubble that is present at this location, or to the possible presence of a stellar bar or misaligned stellar component (see discussion in \S\ref{sec:data} and \S\ref{sec:datacubes}); beyond $\sim 2.5$\,kpc the rotation curve and HI column density map are quite smooth. Since our simple rotation curve model is not able to capture such complexities, we exclude the inner region ($< 2.5$\,kpc) from our fits. Nonetheless, our simple model with just three free parameters ($M_{200}$; $c$ and the disc stellar mass $M_*$) gives a remarkable match to the data over the full rotation curve from $0-5$\,kpc. We obtain tight constraints on $M_{200}$ and a concentration parameter in excellent agreement with cosmological expectations (see the vertical green line in the rightmost panel that marks $c_{\Lambda{\rm CDM}}$ derived from the median $M_{200}$ using the relation in \citealt{2007MNRAS.378...55M}). The priors used for this fit and our derived values for $M_{200}$ and $c$ with 68\% confidence intervals are reported in Table \ref{tab:data}. As in Figure \ref{fig:mock_rotcurve}, the magenta lines overlaid on the rotation curve show what our model rotation curves look like if we switch off cusp-core transformations (i.e. if we input our fitted \coreNFW\ parameters to the NFW profile and calculate the resulting rotation curves). Notice that the dark matter core in NGC 6822 affects the rotation curve out to $\sim 2 R_{1/2}$ (vertical dashed green line), as expected from equation \ref{eqn:coreNFWrotratio} and our discussion in \S\ref{sec:massmodel}.

Similarly to NGC 6822, our model gives an excellent match to WLM's rotation curve over its full range from $0-3$\,kpc. We obtain good constraints on $M_{200}$ and -- like NGC 6822 -- a concentration $c$ consistent with cosmological expectations. As with NGC 6822, it is important that we fit with our \coreNFW\ profile rather than the NFW profile, since the latter leads to rotation curves that rise substantially more steeply than the data (compare the magenta and black lines in the rotation curve data panel).

Both NGC 6822 and WLM have rather low best-fitting $\chi^2_{\rm red}$ (see Table \ref{tab:data}), with $\chi^2_{\rm red} = 0.36$ and $\chi^2_{\rm red} = 0.26$, respectively. Such low values are typical of rotation curve fits and owe in part to the way in which the error bars and derived, and in part -- for NGC 6822 -- to the limited radial range over which the data are fit \citep[see e.g.][]{2010MNRAS.404.1733S,2013MNRAS.433.2314H}. Indeed, if we exclude the `dip' region inside 0.5\,kpc for our mock dwarf from \S\ref{sec:datacubes} (Figure \ref{fig:mock_inclination}, top middle panel), then we find a best-fit $\chi^2_{\rm red} = 0.37$, similar to that for NGC 6822.

\subsubsection{IC 1613: a near face-on disequilibrium dwarf}\label{sec:IC1613}

In contrast to our excellent fits for WLM and NGC 6822, IC 1613 gives a very poor fit (compare the black lines with the red data points in the rotation curve panel). If fitting over the full data range, our model pushes on the lower bound of our $M_{200}$ prior since it cannot reproduce the extremely shallow rise of IC 1613's rotation curve. For this reason we exclude the inner region, fitting over the range $[1.9,2.5]$\,kpc. (This explains why the reduced $\chi_{\rm red}^2 = 0.13$ reported in Table \ref{tab:data} seems surprisingly good despite the obviously poor fit.)

The poor fit for IC 1613 is perhaps not surprising when considering its gas morphology. IC 1613 shows substantial HI bubbles in its HI column density map (leftmost panel) that are qualitatively very similar to the starburst simulation in Figure \ref{fig:mock_rotcurve}; it also presents substantially more recent star formation than all of the other dwarfs considered here, consistent with a starburst (see \S\ref{sec:data}). Most notably, its rotation curve shows prominent wiggles and rises far less steeply than our model would predict given its projected half stellar mass radius.

The shallow rise of IC 1613's rotation curve is reminiscent of our mock starbursting dwarf inclined at $i_{\rm true} = 15^\circ$ that we discussed in \S\ref{sec:datacubes}. There, we showed that for $i_{\rm true} \simlt 30^\circ$, \Barolo\ systematically overestimates the inclination leading to an artificially shallow rise in the rotation curve. The favoured inclination for IC 1613 from \Barolo\ is $i_{\rm fit} = 39 \pm 2^\circ$ (see Table \ref{tab:data}), similar to that derived by Little THINGS \citep{2015AJ....149..180O}. However, our mock data analysis in \S\ref{sec:datacubes} indicates that at this low inclination, the true inclination could be as low as $i_{\rm true} \sim 15^\circ$. In the bottom row of Figure \ref{fig:real_rotcurve}, we consider what the rotation curve for IC 1613 would look like at such an inclination. Now, similarly to our near face-on mock dwarf, the rotation curve rises substantially more steeply. It is well fit by our \coreNFW\ model except at $\sim 0.5$ and $\sim 1.5$\,kpc, where there are prominent HI bubbles. This is similar to the inner region of NGC 6822, where its bubble also appears to cause a depression in the rotation curve, and to our mock rotation curve that we discussed in \S\ref{sec:datacubes}. The mass derived from the rotation curve for IC 1613 is now dramatically different, rising over a dex to $M_{200} = 8_{-2}^{+4}\times 10^9$\,M$_\odot$ (see Table \ref{tab:data}).

With an inclination of $i_{\rm fit} < 40^\circ$, we conclude that it is challenging to recover the true inclination of IC 1613. This is further exacerbated by the fact that it is a starbursting dwarf with prominent fast moving HI bubbles (see \S\ref{sec:data}). We will consider in future work whether improvements to \Barolo\ can yield a more trustworthy measure of its inclination. Until such time, galaxies like IC 1613 are not well-suited for testing $\Lambda$CDM.

\subsubsection{DDO 101: the perils of distance uncertainty}\label{sec:DDO101}

Finally, we consider the interesting and substantially more distant dwarf irregular, DDO 101. The nearest dwarf irregulars to us ($D \simlt 3$\,Mpc) have reliable distances as measured from either Cepheid variable stars or the `tip-of-the-red-giant' branch method \citep[e.g.][]{2012AJ....144....4M}. Very distant dwarfs with $D \simgt 10$\,Mpc enter the Hubble flow and we can also obtain a reliable distance from their redshift alone \citep[e.g.][]{2014MNRAS.443.2204P}. However, for dwarfs at intermediate distances $3 \simlt D \simlt 10$\,Mpc, like DDO 101, we are forced to rely on the `Tully-Fisher' (TF) distance method \citep{1977A&A....54..661T}. This uses the TF relation to obtain the absolute magnitude of the dwarf from its observed peak rotation velocity. Comparing this with the apparent magnitude then allows us to derive its distance. This is problematic for two reasons. Firstly, there is enormous scatter in the TF relation, particularly at low peak rotation velocities \citep[e.g.][]{2016MNRAS.455.3136M}. Secondly, many dwarfs have rising rotation curves and thus their peak rotation velocity is only a {\it lower bound} on the true peak. These difficulties lead to distance errors that can be as large as a factor of two or more. Indeed, the quoted distance range for DDO 101 on the NED database is $6.4 < D_{\rm DDO101}/{\rm Mpc} < 16$. Yet to date, it as been mass modelled assuming $D_{\rm DDO101} = 6.4$\,Mpc \citep{2015AJ....149..180O}.

In Figure \ref{fig:real_rotcurve_ddo101}, we show results from mass modelling DDO 101 assuming three different distances $D_{\rm DDO101} = 6.4$\,Mpc (as assumed previously in the literature; top row); $D_{\rm DDO101} = 12.9$\,Mpc (the Hubble flow distance, using the cosmological parameters from \citealt{2013arXiv1303.5076P}; middle row); and $D_{\rm DDO101} = 16$\,Mpc (the maximum distance reported in NED; bottom row). For $D_{\rm DDO101} = 6.4$\,Mpc, the fit is poor. The rotation curve rises much more steeply than is allowed by the \coreNFW\ model, similarly to the findings in \citet{2015AJ....149..180O}. Indeed, the fact that DDO 101 is one of the few isolated dwarf irregulars with a rotation curve that appears `NFW-like' was the key motivation for including it in our sample here. However, if we place DDO 101 at a larger distance -- as preferred by its redshift -- then the rotation curve is stretched and rises much less steeply. The middle and bottom rows of Figure \ref{fig:real_rotcurve_ddo101} show our results assuming $D_{\rm DDO101} = 12.9$\,Mpc and $D_{\rm DDO101} = 16$\,Mpc, respectively. In both cases, we obtain an excellent fit to the rotation curve, similarly to NGC 6822 and WLM. Note, however, that unlike those galaxies, DDO 101's rotation curve still rises sufficiently steeply that the favoured concentration parameter $c$ pushes on the upper bound of our prior. We would need a higher resolution HI map of similar quality to WLM, and a more accurate measure of $D_{\rm DDO101}$, to determine whether or not this is something to worry about in the context of $\Lambda$CDM; for now, models with a cosmologically reasonable $c$ are still allowed by the data.

\subsubsection{The Baryonic Tully-Fisher Relation (BTFR)}

\begin{figure}
\begin{center}
\includegraphics[width=0.5\textwidth]{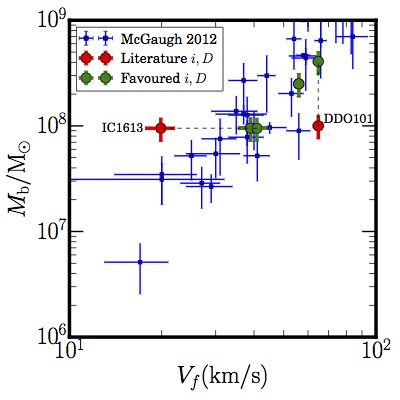}
\caption{The four isolated dIrrs analysed in this paper (red/green data points) as compared to the Baryonic Tully-Fisher Relation (BTFR) of galaxies taken from \citealt{2012AJ....143...40M} (blue data points). Notice that with the literature values, IC 1613 and DDO 101 are outliers from this relation (red data points), but with our favoured inclination and distance of $i_{\rm IC1613} \sim 15^\circ$ and $D_{\rm DDO101} = 12.9$\,Mpc, respectively, they become perfectly consistent (green data points).}
\label{fig:BTFR} 
\end{center}
\end{figure}

The Baryonic Tully-Fisher Relation (BTFR) is a well-known correlation between the total baryonic mass of galaxies (stars plus gas; $M_{\rm b}$) and their peak rotation velocity ($V_f$; e.g. \citealt{2000ApJ...533L..99M,2010ApJ...722..248M,2012AJ....143...40M,2016ApJ...816L..14L}). In Figure \ref{fig:BTFR}, we plot our four isolated dwarfs in the $V_f-M_{\rm b}$ plane (red/green data points) compared to a large sample of gas rich dwarfs taken from \citet{2012AJ....143...40M} (blue data points). Interestingly, our two `good' galaxies -- WLM and NGC 6822 lie within the 1-sigma scatter of the relation (green data points), whereas our two `problem galaxies', IC 1613 and DDO 101 are both significant outliers (red data points). However, if we assume $D_{\rm DDO101} = 12.9$\,Mpc and $i_{\rm IC1613} \sim 15^\circ$ -- as favoured by their \coreNFW\ rotation curve fits -- then both galaxies become consistent with the BTFR (green data points). This is encouraging as it implies that with our favoured distance and inclination for DDO 101 and IC 1613, respectively, these two dIrrs become very similar to other dIrrs of similar baryonic mass. This behaviour was noted also in \citet{2016arXiv160101026O} where they pointed out that for IC 1613 a mean inclination of $i \sim 22^\circ$ would be required to bring it onto the BTFR, similarly to what we find here.

\section{Conclusions} \label{sec:conclusions}

In the first part of this paper, we used mock data from high resolution simulations of isolated dwarf galaxies to understand the shape and diversity of dwarf galaxy rotation curves in $\Lambda$CDM. Our key findings were as follows:

\begin{itemize}

\item The rotation curve in our mock dwarfs systematically shifts up and down in amplitude throughout the starburst cycle. At its most quiescent phase, the asymmetric drift corrected gas rotational velocities give a good match to the true circular speed curve. Following a strong starburst, however, the rotation curve rises more steeply, while just post-starburst it rises substantially less steeply. Such `disequilibrium' galaxies can be readily identified, however, by the presence of substantial fast-moving HI bubbles (with expansion velocity $\sim 10-30$\,km/s). Our simulated dwarfs continually cycle through quiescent, starburst and post-starburst phases suggesting that all three situations should be common. Indeed, many real starbursting dwarfs show kinematically disturbed HI discs \citep[e.g.][]{2014A&A...566A..71L}.

\item Fitting a new \coreNFW\ profile (that accounts for dark matter cusp-core transformations due to stellar feedback) to the above mock rotation curve data, we found that we could successfully recover the virial mass $M_{200}$ and concentration parameter $c$ of our mock data, but only if we fit the rotation curve in its quiescent phase. The starburst and post-starburst mocks led to a systematic under-estimate of $M_{200}$ by up to half a dex, and a biased concentration parameter $c$.

\item It is important to use the \coreNFW\ profile rather than an NFW profile in the mass models. The NFW profile rotation curves rise too steeply to be consistent with our mock data and lead to a substantial bias on $M_{200}$ and/or $c$ if used instead of the \coreNFW\ profile.

\item We tested our recovery of the rotation curve from mock HI datacubes, using the \Barolo\ code. We found that we could obtain an excellent recovery of the rotation curve, provided its best fit inclination was $i_{\rm fit} \simgt 40^\circ$. Lower inclinations than this are systematically biased high, leading to rotation curves that have an artificially shallow rise, with an associated underestimate of $M_{200}$ of up to a dex or more.

\item The stellar kinematics are substantially less affected by the starburst cycle than the gas kinematics. This might suggest that for starburst or post-starburst systems, stars will be a more reliable probe of the underlying potential than the HI rotation curve. However, stars probe the potential only out to $R_{1/2}$, making them a poor probe of the halo virial mass $M_{200}$, as discussed in R16. Thus, even when including stellar kinematic data, it is probably best to avoid galaxies that show signs of having a gaseous rotation curve that is far from equilibrium. 

\end{itemize} 
\noindent
We then went on to fit mass models, using a new \coreNFW\ profile that accounts for cusp-core transformations due to stellar feedback, to real data for four isolated dwarf irregular galaxies. These were chosen to span an interesting range of rotation curve morphologies: NGC 6822; WLM; IC 1613; and DDO 101. Our key results were as follows: 

\begin{itemize}

\item We obtained excellent fits for NGC 6822 and WLM, with tight constraints on $M_{200}$, and concentration parameters $c$ consistent with cosmological expectations (see Table \ref{tab:data} and Figure \ref{fig:real_rotcurve}).

\item By contrast, both IC 1613 and DDO 101 gave a very poor fit. For IC 1613, we showed that this owes to substantial fast moving HI bubbles and a poorly determined inclination; for DDO 101, the problem was its uncertain distance. Using $i_{\rm IC1613} \sim 15^\circ$ and $D_{\rm DDO101} \sim 12$\,Mpc, consistent with current uncertainties, we were able to fit both galaxies very well (see Table \ref{tab:data} and Figures \ref{fig:real_rotcurve} and \ref{fig:real_rotcurve_ddo101}). Interestingly, with this inclination and distance, IC 1613 and DDO 101 also become consistent with the `Baryonic Tully Fisher Relation' of galaxies (Figure \ref{fig:BTFR}).

\end{itemize} 
Although in this paper we have only analysed a small sample of four dwarf irregular galaxies, we have deliberately picked some of the most challenging rotation curves for $\Lambda$CDM reported in the literature to date. It is encouraging, then, that all four are well-fit by our \coreNFW\ model once errors in the inclination and distance are properly taken into account. This suggests that the now long-standing `cusp-core' problem owes to previously unmodelled `baryonic' physics (bursty star formation driven by stellar feedback), while the newer `dwarf rotation curve diversity' problem likely owes to a mixture of inclination and distance error, and disequilibria driven by fast moving HI bubbles. There are, however, other sources of rotation curve scatter that we have not considered here, including cosmological scatter on the \coreNFW\ model parameters that we only briefly explored in Appendix \ref{app:etafree}, and the effect of galaxy mergers and interactions. We will consider the effect of these, and confront our \coreNFW\ model with a larger sample of dwarf irregulars in forthcoming papers.

\section{Acknowledgements}

JIR and OA would like to acknowledge support from STFC consolidated grant ST/M000990/1. JIR acknowledges support from the MERAC foundation. OA would like to acknowledge support from the Swedish Research Council (grant 2014-5791). This research made use of APLpy, an open-source plotting package for Python hosted at \href{http://aplpy.github.com}{http://aplpy.github.com}; PyNbody for the simulation analysis (\href{https://github.com/pynbody/pynbody}{https://github.com/pynbody/pynbody}; \citealt{2013ascl.soft05002P}); and the \EMCEE\ package for model fitting (\href{http://dan.iel.fm/emcee/current/}{http://dan.iel.fm/emcee/current/}). All simulations were run on the Surrey Galaxy Factory. The research has made use of the NASA/IPAC Extragalactic Database (NED) which is operated by the Jet Propulsion Laboratory, California Institute of Technology, under contract with the National Aeronautics and Space Administration. We would like to thank Se-Heon Oh and Little THINGS for kindly making their data public; and Erwin de Blok for kindly providing the data for NGC 6822. We would like to thank Jorge Pe\~narrubia; Octavio Valenzuela; Malcolm Fairbairn; Chris Brook; Benoit Famaey and the anonymous referee for useful comments that improved this manuscript.

\appendix

\section{The effect of varying the asymmetric drift correction}\label{app:asymcorrect} 

\begin{figure}
\begin{center}
\includegraphics[width=0.5\textwidth]{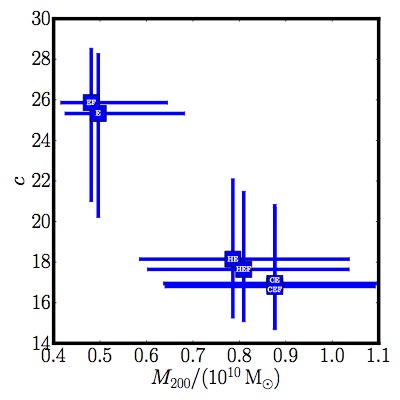}
\caption{The effect of six different asymmetric drift corrections on the derived $M_{200}$ and $c$ for WLM. The corrections are described in \S\ref{app:asymcorrect} and marked at the centre of each data point. Notice that in all cases, the derived $M_{200}$ and $c$ overlap within our quoted uncertainties.}
\label{fig:ad_fix_compare} 
\end{center}
\end{figure}

In this appendix, we explore the effect of six different asymmetric drift (AD) corrections on the derived rotation curve for WLM. In each case, we solve equation \ref{eqn:sigmaD} assuming different radially varying gas surface density profiles $\Sigma_{\rm gas}(R)$ and assuming either a fixed gas velocity dispersion, or one that is also radially varying (as measured from the data):

\begin{itemize} 
\item {\bf EF:} This is an exponential, constant velocity dispersion, AD correction described already in \S\ref{sec:asymcorrect}.
\item {\bf E:} The is an exponential AD correction, but using the measured radially varying velocity dispersion.
\item {\bf CEF:} This is a core+exponential AD correction that assumes the following functional form: 
\begin{equation} 
\Sigma_{\rm gas} \sigma_{\rm gas}^2 = \frac{I_0 (R_0 + 1)}{R_0 + e^{\alpha R}}
\end{equation}
where $I_0, R_0$ and $\alpha$ are fitting parameters, and we assume a fixed velocity dispersion $\sigma_{\rm gas} = {\rm const.}$ 
\item {\bf CE:} This is a core+exponential AD correction with radially varying velocity dispersion. 
\item {\bf HEF:} This is a hole+exponential AD correction:
\begin{equation} 
\Sigma_{\rm gas} \sigma_{\rm gas}^2 = I_0 \left(1 + \frac{R}{R_0}\right)^{\alpha} e^{-\frac{R}{R_0}}
\end{equation}
with fixed velocity dispersion.
\item {\bf HEF:} This is a hole+exponential AD correction with radially varying velocity dispersion. 
\end{itemize}
The derived rotation curve for WLM in each case agrees within our 1-$\sigma$ uncertainties. In Figure \ref{fig:ad_fix_compare}, we show the $M_{200}$ and $c$ derived from these rotation curves, with 68\% confidence intervals marked for each case (each data point is labelled with its correction). Notice that the error bars for both $M_{200}$ and $c$ overlap in all cases. There is a systematic trend for the exponential correction to systematically favour lower $M_{200}$ and higher $c$, but it is not statistically significant. For the mock data, we use the EF correction since an exponential gives a good fit to the gas surface density profile while the velocity dispersion profile is found to be very close to constant; for the real data we favour the CE correction since this gives a slightly better representation of the HI data. The effect of these choices, however, is very small.

\section{Fitting NFW instead of \bfcoreNFW}\label{app:nfwfit}

\begin{figure*}
\begin{center}
\includegraphics[width=0.99\textwidth]{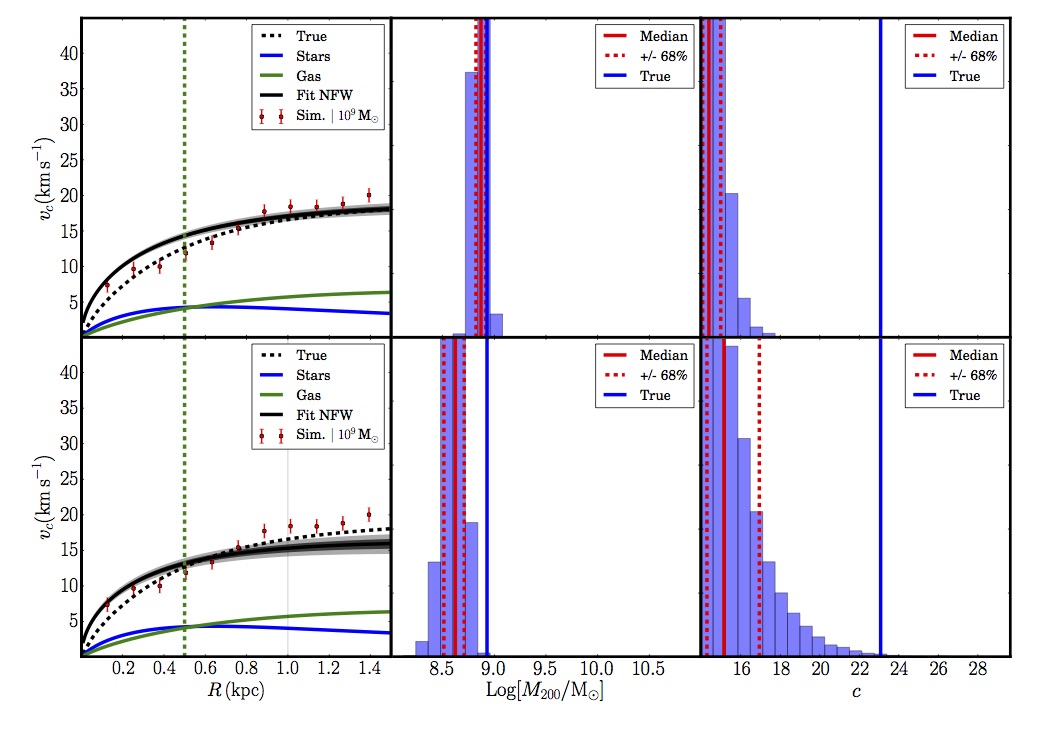}
\caption{The effect of fitting the NFW profile instead of \coreNFW\ to our mock data. The lines and symbols are as in Figure \ref{fig:mock_rotcurve}. The top row shows a fit to the mock rotation curve for our quiescent $M_{200} = 10^9$\,M$_\odot$ dwarf. Since the data reach to $R = 1.5$\,kpc where the rotation curve is flat, we still recover the correct $M_{200}$. The concentration parameter is, however, biased towards low $c$ and pushes on our prior. The bottom row shows the same fit but excluding data for $R > 1$\,kpc. Now in order to fit the shallow rise of the rotation curve, the NFW profile is pushed towards systematically low $M_{200}$.}
\label{fig:mock_nfwfit} 
\end{center}
\end{figure*}

In this appendix, we explore the effect of fitting an NFW profile to our mock data instead of our \coreNFW\ profile (see \S\ref{sec:rotmethod} for a description of both profiles). In Figure \ref{fig:mock_nfwfit}, we show fits to our mock rotation curve for our quiescent $M_{200} = 10^9$\,M$_\odot$ dwarf, but assuming an NFW profile for the dark matter halo. The top row shows a fit using all of the data to $R = 1.5$\,kpc. Since this reaches out to where the rotation curve is flat, we still recover the correct $M_{200}$. The concentration parameter is, however, biased towards low $c$ and pushes on our theory prior. The bottom row shows the same fit but excluding data for $R > 1$\,kpc. Now, in order to fit the shallow rise of the rotation curve, the NFW profile is pushed towards systematically low $M_{200}$.

\section{Allowing the dark matter core size to vary}\label{app:etafree} 

\begin{figure}
\begin{center}
\includegraphics[width=0.49\textwidth]{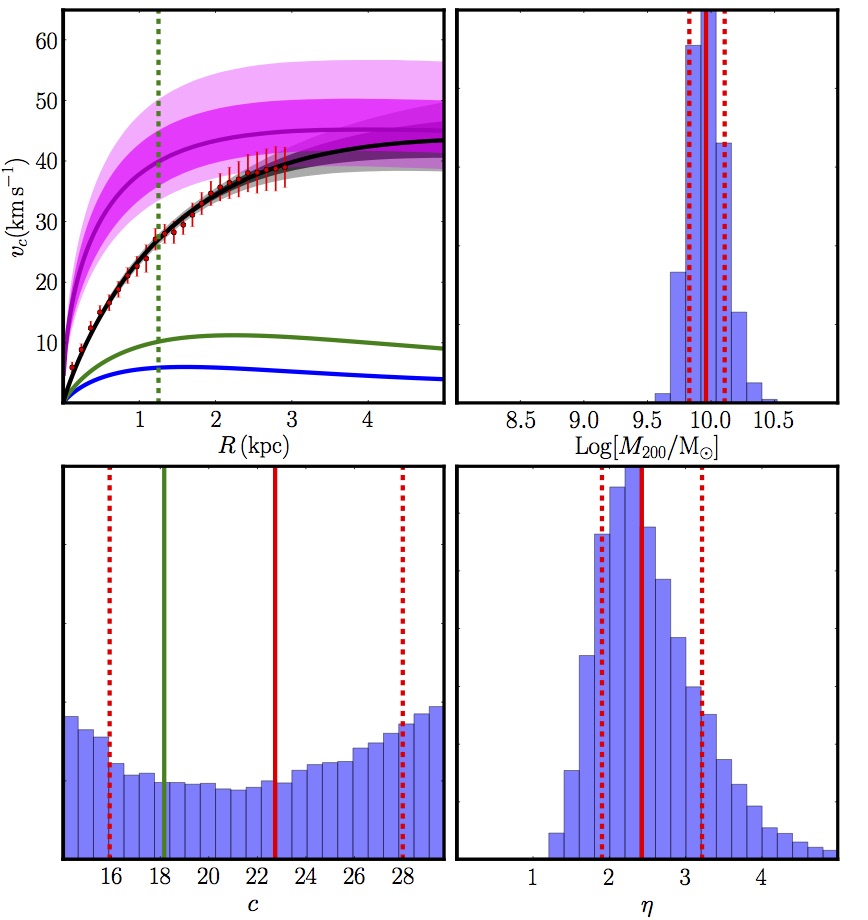}
\caption{The effect of allowing the dark matter core size to freely vary when fitting data for WLM. The lines and symbols are as in Figure \ref{fig:mock_rotcurve}. Recall that the dark matter core size is set by equation \ref{eqn:etarc}: $r_c = \eta R_{1/2}$ and is therefore controlled by the dimensionless parameter $\eta$. In this fit to WLM's rotation curve, we allow $0 < \eta < 5$; the histogram of recovered $\eta$ from our model chains is shown in the bottom right panel. Notice that the data clearly favour a dark matter core over a cusp ($\eta = 0$), while our favoured value of $\eta = 1.75$ from R16 is recovered at the edge of our 68\% confidence intervals.}
\label{fig:etafree} 
\end{center}
\end{figure}

By default, we have assumed throughout this paper that the dark matter core size parameter $\eta = 1.75$ (equation \ref{eqn:etarc}) is held fixed. This gave the best match to our simulations in R16. However, as discussed in R16, there could be some scatter on $\eta$ due to varying halo spin, concentration parameter and/or halo assembly history. For this reason, in this Appendix we explore allowing $\eta$ to vary freely over the range $0 < \eta < 5$ when fitting data for WLM. The results are shown in Figure \ref{fig:etafree}. As might be expected, allowing $\eta$ to vary limits our ability to measure the halo concentration parameter $c$ and slightly inflates our errors on $M_{200}$. Otherwise, however, its effect is benign. Interestingly, we find $\eta_{\rm WLM} = 2.4_{-0.52}^{+0.78}$ at 68\% confidence, consistent with our favoured $\eta = 1.75$ and clearly inconsistent with a dark matter cusp ($\eta = 0$).

\section{The effect of the concentration parameter prior}\label{app:cprior} 

In this appendix, we explore the effect of changing our concentration parameter prior on the rotation curve fits and $M_{200}$ for WLM. We compare our default flat linear prior: $14 < c < 30$ with a generous logarithmic prior: $0.9 < \ln c < 1.6$ and a `cosmology prior'. This latter is the most restrictive prior on $c$ that we consider. It fixes the concentration parameter using the mass-concentration relation from \citet{2007MNRAS.378...55M}:

\begin{equation}
\log_{10} c = 1.02 - 0.109 \left(\log_{10}\left(\frac{M_{200}}{{\rm M}_\odot} h\right)-12\right)
\label{eqn:conc_mass}
\end{equation}  
This same relation is used to set the upper and lower bounds on our default and logarithmic priors. For our default prior, the lower/upper bound is set by cosmic mean $c$ for halos of mass $M_{200} = 10^8$\,M$_\odot$ and $M_{200} = 10^{11}$\,M$_\odot$, respectively. This  comfortably brackets the expected range of halo masses for the low mass dwarf irregulars that we consider in this work. The logarithmic prior is even more generous, with its boundaries set by halos of mass $M_{200} = 10^7$\,M$_\odot$ and $M_{200} = 10^{13}$\,M$_\odot$, respectively. Thus, the logarithmic prior is deliberately allowed to explore concentration parameters that are unrealistically low and high as compared to real halos in $\Lambda$CDM. This allows us to asses how important priors on $c$ are for our analysis. 

\begin{figure}
\begin{center}
\includegraphics[width=0.49\textwidth]{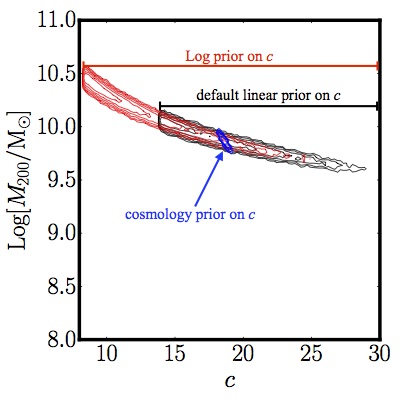}
\caption{The effect of varying the prior on the concentration parameter $c$ for WLM. The red contours show the density of our MCMC models ($\rho_{\rm models}$) in the $c-M_{200}$ plane for our most generous logarithmic prior (that allows $c$ to extend to cosmologically inconsistent values); the black contours show the same for our already generous default prior; and the blue contours show results for a very restrictive `cosmology' prior that demands that $c$ exactly obeys the concentration-mass relation in equation \ref{eqn:conc_mass}. In all cases, the contours mark ten logarithmically spaced isodensity contours of the MCMC models over the range $0.01 < \rho_{\rm models} < 1$, where $\rho_{\rm models}$ is normalised such that $\max(\rho_{\rm models}) = 1$.}
\label{fig:cprior} 
\end{center}
\end{figure}

Figure \ref{fig:cprior} shows the density of our MCMC models ($\rho_{\rm models}$) projected onto the $c-M_{200}$ plane for WLM. The red contours show the results for our most generous logarithmic prior that allows $c$ to extend to cosmologically inconsistent values; the black contours show the same for our already generous default prior; and the blue contours show results for a very restrictive `cosmology' prior that demands that $c$ exactly obeys the concentration-mass relation in equation \ref{eqn:conc_mass}. In all cases, the contours mark ten logarithmically spaced isodensity contours of the MCMC models over the range $0.01 < \rho_{\rm models} < 1$, where $\rho_{\rm models}$ is normalised such that $\max(\rho_{\rm models}) = 1$. As can be seen, for the default and logarithmic prior there is a `banana' degeneracy between $M_{200}$ and $c$, while models with low $c$ are weakly favoured. This trend to low $c$ is, however, not statistically significant. The best $\chi^2_{\rm red}$ for the restrictive cosmology prior was $\chi^2_{\rm red} = 0.1$ as compared to $\chi^2_{\rm red} = 0.08$ for the logarithmic prior. Rather, this behaviour simply indicates that the data are not able to give tight constraints on $c$ and thus the marginalised distribution of $c$ depends wholly on our priors.

Finally, we may reasonably ask if the higher $M_{200}$ allowed by the logarithmic prior are acceptable models for WLM. In the sense that they fit the rotation curve data, they are acceptable models. However, such halos lie far from the cosmic mean concentration parameter and so are not expected in $\Lambda$CDM. (From equation \ref{eqn:conc_mass}, we expect a halo of mass $\log_{10} M_{200} = 10.5$ to have $c_{10.5} \sim 16$.) Thus, our default prior on $c$ is a compromise between allowing $c$ to vary over a generous range while disallowing halos that are unlikely to exist in $\Lambda$CDM.

\bibliographystyle{mn2e}
\bibliography{refs,ref2}

\begin{thebibliography}{133}
\expandafter\ifx\csname natexlab\endcsname\relax\def\natexlab#1{#1}\fi

\bibitem[{{Adams} {et~al}\mbox{.}(2014){Adams}, {Simon}, {Fabricius}, {van den
  Bosch}, {Barentine}, {Bender}, {Gebhardt}, {Hill}, {Murphy}, {Swaters},
  {Thomas}, \& {van de Ven}}]{2014ApJ...789...63A}
{Adams} J.~J. {et~al.}, 2014, \apj, 789, 63

\bibitem[{{Agertz} \& {Kravtsov}(2015)}]{2015ApJ...804...18A}
{Agertz} O., {Kravtsov} A.~V., 2015, \apj, 804, 18

\bibitem[{{Agertz} {et~al}\mbox{.}(2013){Agertz}, {Kravtsov}, {Leitner}, \&
  {Gnedin}}]{2013ApJ...770...25A}
{Agertz} O., {Kravtsov} A.~V., {Leitner} S.~N., {Gnedin} N.~Y., 2013, \apj,
  770, 25

\bibitem[{{Alcubierre} {et~al}\mbox{.}(2002){Alcubierre}, {Guzm{\'a}n},
  {Matos}, {N{\'u}{\~n}ez}, {Ure{\~n}a-L{\'o}pez}, \&
  {Wiederhold}}]{2002CQGra..19.5017A}
{Alcubierre} M., {Guzm{\'a}n} F.~S., {Matos} T., {N{\'u}{\~n}ez} D.,
  {Ure{\~n}a-L{\'o}pez} L.~A., {Wiederhold} P., 2002, Classical and Quantum
  Gravity, 19, 5017

\bibitem[{{Avila-Reese} {et~al}\mbox{.}(2001){Avila-Reese}, {Col{\'{\i}}n},
  {Valenzuela}, {D'Onghia}, \& {Firmani}}]{2001ApJ...559..516A}
{Avila-Reese} V., {Col{\'{\i}}n} P., {Valenzuela} O., {D'Onghia} E., {Firmani}
  C., 2001, \apj, 559, 516

\bibitem[{Baade(1929)}]{1929AN....234..407B}
Baade W., 1929, Astronomical news, 234, 407

\bibitem[{{Barnard}(1884)}]{1884AN....110..125B}
{Barnard} E.~E., 1884, Astronomische Nachrichten, 110, 125

\bibitem[{{Bekenstein}(2004)}]{2004PhRvD..70h3509B}
{Bekenstein} J.~D., 2004, \prd, 70, 083509

\bibitem[{{Bekki} \& {Chiba}(2006)}]{2006ApJ...637L..97B}
{Bekki} K., {Chiba} M., 2006, \apjl, 637, L97

\bibitem[{{Bertone}(2010)}]{2010pdmo.book.....B}
{Bertone} G., 2010, {Particle Dark Matter : Observations, Models and Searches}.
  Cambridge University Press

\bibitem[{{Bigiel} {et~al}\mbox{.}(2011){Bigiel}, {Leroy}, {Walter}, {Brinks},
  {de Blok}, {Kramer}, {Rix}, {Schruba}, {Schuster}, {Usero}, \&
  {Wiesemeyer}}]{Bigiel2011}
{Bigiel} F. {et~al.}, 2011, \apjl, 730, L13

\bibitem[{{Binney} \& {Tremaine}(2008)}]{1987gady.book.....B}
{Binney} J., {Tremaine} S., 2008, {Galactic dynamics}. Princeton, NJ, Princeton
  University Press, 2008, 747 p.

\bibitem[{{Bode} {et~al}\mbox{.}(2001){Bode}, {Ostriker}, \&
  {Turok}}]{2001ApJ...556...93B}
{Bode} P., {Ostriker} J.~P., {Turok} N., 2001, \apj, 556, 93

\bibitem[{{Boley} {et~al}\mbox{.}(2009){Boley}, {Lake}, {Read}, \&
  {Teyssier}}]{2009arXiv0908.1254B}
{Boley} A.~C., {Lake} G., {Read} J., {Teyssier} R., 2009, ArXiv e-prints

\bibitem[{{Boyarsky} {et~al}\mbox{.}(2009){Boyarsky}, {Ruchayskiy}, \&
  {Shaposhnikov}}]{2009ARNPS..59..191B}
{Boyarsky} A., {Ruchayskiy} O., {Shaposhnikov} M., 2009, Annual Review of
  Nuclear and Particle Science, 59, 191

\bibitem[{{Brook}(2015)}]{2015MNRAS.454.1719B}
{Brook} C.~B., 2015, \mnras, 454, 1719

\bibitem[{{Bruhweiler} {et~al}\mbox{.}(1980){Bruhweiler}, {Gull}, {Kafatos}, \&
  {Sofia}}]{1980ApJ...238L..27B}
{Bruhweiler} F.~C., {Gull} T.~R., {Kafatos} M., {Sofia} S., 1980, \apjl, 238,
  L27

\bibitem[{{Bullock} {et~al}\mbox{.}(2001){Bullock}, {Dekel}, {Kolatt},
  {Kravtsov}, {Klypin}, {Porciani}, \& {Primack}}]{BullockEtal2001b}
{Bullock} J.~S., {Dekel} A., {Kolatt} T.~S., {Kravtsov} A.~V., {Klypin} A.~A.,
  {Porciani} C., {Primack} J.~R., 2001, \apj, 555, 240

\bibitem[{{Campbell} {et~al}\mbox{.}(2016){Campbell}, {Frenk}, {Jenkins},
  {Eke}, {Navarro}, {Sawala}, {Schaller}, {Fattahi}, {Oman}, \&
  {Theuns}}]{2016arXiv160304443C}
{Campbell} D.~J.~R. {et~al.}, 2016, ArXiv e-prints

\bibitem[{{Cannon} {et~al}\mbox{.}(2012){Cannon}, {O'Leary}, {Weisz},
  {Skillman}, {Dolphin}, {Bigiel}, {Cole}, {de Blok}, \&
  {Walter}}]{2012ApJ...747..122C}
{Cannon} J.~M. {et~al.}, 2012, \apj, 747, 122

\bibitem[{{Clowe} {et~al}\mbox{.}(2006){Clowe}, {Brada{\v c}}, {Gonzalez},
  {Markevitch}, {Randall}, {Jones}, \& {Zaritsky}}]{2006ApJ...648L.109C}
{Clowe} D., {Brada{\v c}} M., {Gonzalez} A.~H., {Markevitch} M., {Randall}
  S.~W., {Jones} C., {Zaritsky} D., 2006, \apjl, 648, L109

\bibitem[{{de Blok}(2009)}]{2009arXiv0910.3538D}
{de Blok} W.~J.~G., 2009, ArXiv e-prints

\bibitem[{{de Blok} {et~al}\mbox{.}(2001){de Blok}, {McGaugh}, {Bosma}, \&
  {Rubin}}]{2001ApJ...552L..23D}
{de Blok} W.~J.~G., {McGaugh} S.~S., {Bosma} A., {Rubin} V.~C., 2001, \apjl,
  552, L23

\bibitem[{{de Blok} \& {Walter}(2000)}]{2000ApJ...537L..95D}
{de Blok} W.~J.~G., {Walter} F., 2000, \apjl, 537, L95

\bibitem[{{de Blok} {et~al}\mbox{.}(2008){de Blok}, {Walter}, {Brinks},
  {Trachternach}, {Oh}, \& {Kennicutt}}]{2008AJ....136.2648D}
{de Blok} W.~J.~G., {Walter} F., {Brinks} E., {Trachternach} C., {Oh} S.-H.,
  {Kennicutt}, Jr. R.~C., 2008, \aj, 136, 2648

\bibitem[{{Demers} {et~al}\mbox{.}(2006){Demers}, {Battinelli}, \&
  {Kunkel}}]{2006ApJ...636L..85D}
{Demers} S., {Battinelli} P., {Kunkel} W.~E., 2006, \apjl, 636, L85

\bibitem[{{Di Cintio} {et~al}\mbox{.}(2014){Di Cintio}, {Brook}, {Dutton},
  {Macci{\`o}}, {Stinson}, \& {Knebe}}]{2014MNRAS.441.2986D}
{Di Cintio} A., {Brook} C.~B., {Dutton} A.~A., {Macci{\`o}} A.~V., {Stinson}
  G.~S., {Knebe} A., 2014, \mnras, 441, 2986

\bibitem[{{Di Teodoro} \& {Fraternali}(2015)}]{2015MNRAS.451.3021D}
{Di Teodoro} E.~M., {Fraternali} F., 2015, \mnras, 451, 3021

\bibitem[{{Dodelson}(2011)}]{2011IJMPD..20.2749D}
{Dodelson} S., 2011, International Journal of Modern Physics D, 20, 2749

\bibitem[{{Dubinski} \& {Carlberg}(1991)}]{1991ApJ...378..496D}
{Dubinski} J., {Carlberg} R.~G., 1991, \apj, 378, 496

\bibitem[{{Dubois} \& {Teyssier}(2008)}]{dubois08}
{Dubois} Y., {Teyssier} R., 2008, \aap, 477, 79

\bibitem[{{El-Badry} {et~al}\mbox{.}(2015){El-Badry}, {Wetzel}, {Geha},
  {Hopkins}, {Kere{\v s}}, {Chan}, \&
  {Faucher-Gigu{\`e}re}}]{2015arXiv151201235E}
{El-Badry} K., {Wetzel} A.~R., {Geha} M., {Hopkins} P.~F., {Kere{\v s}} D.,
  {Chan} T.~K., {Faucher-Gigu{\`e}re} C.-A., 2015, ArXiv e-prints

\bibitem[{{Elbert} {et~al}\mbox{.}(2014){Elbert}, {Bullock}, {Garrison-Kimmel},
  {Rocha}, {O{\~n}orbe}, \& {Peter}}]{2014arXiv1412.1477E}
{Elbert} O.~D., {Bullock} J.~S., {Garrison-Kimmel} S., {Rocha} M., {O{\~n}orbe}
  J., {Peter} A.~H.~G., 2014, ArXiv e-prints

\bibitem[{{Flores} \& {Primack}(1994)}]{1994ApJ...427L...1F}
{Flores} R.~A., {Primack} J.~R., 1994, \apjl, 427, L1

\bibitem[{{Foreman-Mackey} {et~al}\mbox{.}(2013){Foreman-Mackey}, {Hogg},
  {Lang}, \& {Goodman}}]{2013PASP..125..306F}
{Foreman-Mackey} D., {Hogg} D.~W., {Lang} D., {Goodman} J., 2013, \pasp, 125,
  306

\bibitem[{{Freeman}(1970)}]{Freeman1970}
{Freeman} K.~C., 1970, \apj, 160, 811

\bibitem[{{Gnedin} \& {Zhao}(2002)}]{2002MNRAS.333..299G}
{Gnedin} O.~Y., {Zhao} H., 2002, \mnras, 333, 299

\bibitem[{{Goerdt} {et~al}\mbox{.}(2006){Goerdt}, {Moore}, {Read}, {Stadel}, \&
  {Zemp}}]{Goerdt:2006rw}
{Goerdt} T., {Moore} B., {Read} J.~I., {Stadel} J., {Zemp} M., 2006, \mnras,
  368, 1073

\bibitem[{{G{\'o}rski} {et~al}\mbox{.}(2011){G{\'o}rski}, {Pietrzy{\'n}ski}, \&
  {Gieren}}]{2011AJ....141..194G}
{G{\'o}rski} M., {Pietrzy{\'n}ski} G., {Gieren} W., 2011, \aj, 141, 194

\bibitem[{{Governato} {et~al}\mbox{.}(2010){Governato}, {Brook}, {Mayer},
  {Brooks}, {Rhee}, {Wadsley}, {Jonsson}, {Willman}, {Stinson}, {Quinn}, \&
  {Madau}}]{GovernatoEtAl2010}
{Governato} F. {et~al.}, 2010, \nat, 463, 203

\bibitem[{{Hague} \& {Wilkinson}(2013)}]{2013MNRAS.433.2314H}
{Hague} P.~R., {Wilkinson} M.~I., 2013, \mnras, 433, 2314

\bibitem[{{Hague} \& {Wilkinson}(2014)}]{2014MNRAS.443.3712H}
{Hague} P.~R., {Wilkinson} M.~I., 2014, \mnras, 443, 3712

\bibitem[{{Hayashi} {et~al}\mbox{.}(2004){Hayashi}, {Navarro}, {Power},
  {Jenkins}, {Frenk}, {White}, {Springel}, {Stadel}, \&
  {Quinn}}]{2004MNRAS.355..794H}
{Hayashi} E. {et~al.}, 2004, \mnras, 355, 794

\bibitem[{{Hopkins} {et~al}\mbox{.}(2013){Hopkins}, {Narayanan}, \&
  {Murray}}]{2013MNRAS.432.2647H}
{Hopkins} P.~F., {Narayanan} D., {Murray} N., 2013, \mnras, 432, 2647

\bibitem[{{Hopkins} {et~al}\mbox{.}(2011){Hopkins}, {Quataert}, \&
  {Murray}}]{2011MNRAS.417..950H}
{Hopkins} P.~F., {Quataert} E., {Murray} N., 2011, \mnras, 417, 950

\bibitem[{{Hunter} {et~al}\mbox{.}(2012){Hunter}, {Ficut-Vicas}, {Ashley},
  {Brinks}, {Cigan}, {Elmegreen}, {Heesen}, {Herrmann}, {Johnson}, {Oh},
  {Rupen}, {Schruba}, {Simpson}, {Walter}, {Westpfahl}, {Young}, \&
  {Zhang}}]{2012AJ....144..134H}
{Hunter} D.~A. {et~al.}, 2012, \aj, 144, 134

\bibitem[{{Irwin} {et~al}\mbox{.}(2007){Irwin}, {Belokurov}, {Evans},
  {Ryan-Weber}, {de Jong}, {Koposov}, {Zucker}, {Hodgkin}, {Gilmore}, {Prema},
  {Hebb}, {Begum}, {Fellhauer}, {Hewett}, {Kennicutt}, {Wilkinson}, {Bramich},
  {Vidrih}, {Rix}, {Beers}, {Barentine}, {Brewington}, {Harvanek},
  {Krzesinski}, {Long}, {Nitta}, \& {Snedden}}]{2007ApJ...656L..13I}
{Irwin} M.~J. {et~al.}, 2007, \apjl, 656, L13

\bibitem[{{Jungman} {et~al}\mbox{.}(1996){Jungman}, {Kamionkowski}, \&
  {Griest}}]{1996PhR...267..195J}
{Jungman} G., {Kamionkowski} M., {Griest} K., 1996, \physrep, 267, 195

\bibitem[{{Kannan} {et~al}\mbox{.}(2012){Kannan}, {Macci{\`o}}, {Pasquali},
  {Moster}, \& {Walter}}]{2012ApJ...746...10K}
{Kannan} R., {Macci{\`o}} A.~V., {Pasquali} A., {Moster} B.~P., {Walter} F.,
  2012, \apj, 746, 10

\bibitem[{{Karlsson} {et~al}\mbox{.}(2013){Karlsson}, {Bromm}, \&
  {Bland-Hawthorn}}]{2013RvMP...85..809K}
{Karlsson} T., {Bromm} V., {Bland-Hawthorn} J., 2013, Reviews of Modern
  Physics, 85, 809

\bibitem[{{Kauffmann}(2014)}]{2014MNRAS.441.2717K}
{Kauffmann} G., 2014, \mnras, 441, 2717

\bibitem[{{Kepley} {et~al}\mbox{.}(2007){Kepley}, {Wilcots}, {Hunter}, \&
  {Nordgren}}]{2007AJ....133.2242K}
{Kepley} A.~A., {Wilcots} E.~M., {Hunter} D.~A., {Nordgren} T., 2007, \aj, 133,
  2242

\bibitem[{{Kimm} {et~al}\mbox{.}(2015){Kimm}, {Cen}, {Devriendt}, {Dubois}, \&
  {Slyz}}]{2015arXiv150105655K}
{Kimm} T., {Cen} R., {Devriendt} J., {Dubois} Y., {Slyz} A., 2015, ArXiv
  e-prints

\bibitem[{{Kirby} {et~al}\mbox{.}(2014){Kirby}, {Bullock}, {Boylan-Kolchin},
  {Kaplinghat}, \& {Cohen}}]{2014MNRAS.439.1015K}
{Kirby} E.~N., {Bullock} J.~S., {Boylan-Kolchin} M., {Kaplinghat} M., {Cohen}
  J.~G., 2014, \mnras, 439, 1015

\bibitem[{{Kravtsov}(2003)}]{2003ApJ...590L...1K}
{Kravtsov} A.~V., 2003, \apjl, 590, L1

\bibitem[{{Kuzio de Naray} \& {Kaufmann}(2011)}]{2011MNRAS.414.3617K}
{Kuzio de Naray} R., {Kaufmann} T., 2011, \mnras, 414, 3617

\bibitem[{{Lake} \& {Skillman}(1989)}]{1989AJ.....98.1274L}
{Lake} G., {Skillman} E.~D., 1989, \aj, 98, 1274

\bibitem[{{Leaman} {et~al}\mbox{.}(2012){Leaman}, {Venn}, {Brooks},
  {Battaglia}, {Cole}, {Ibata}, {Irwin}, {McConnachie}, {Mendel}, \&
  {Tolstoy}}]{2012ApJ...750...33L}
{Leaman} R. {et~al.}, 2012, \apj, 750, 33

\bibitem[{{Lelli} {et~al}\mbox{.}(2016){Lelli}, {McGaugh}, \&
  {Schombert}}]{2016ApJ...816L..14L}
{Lelli} F., {McGaugh} S.~S., {Schombert} J.~M., 2016, \apjl, 816, L14

\bibitem[{{Lelli} {et~al}\mbox{.}(2014){Lelli}, {Verheijen}, \&
  {Fraternali}}]{2014A&A...566A..71L}
{Lelli} F., {Verheijen} M., {Fraternali} F., 2014, \aap, 566, A71

\bibitem[{{Lozinskaya}(2002)}]{2002A&AT...21..223L}
{Lozinskaya} T.~A., 2002, Astronomical and Astrophysical Transactions, 21, 223

\bibitem[{{Macci{\`o}} {et~al}\mbox{.}(2007){Macci{\`o}}, {Dutton}, {van den
  Bosch}, {Moore}, {Potter}, \& {Stadel}}]{2007MNRAS.378...55M}
{Macci{\`o}} A.~V., {Dutton} A.~A., {van den Bosch} F.~C., {Moore} B., {Potter}
  D., {Stadel} J., 2007, \mnras, 378, 55

\bibitem[{{Macci{\`o}} {et~al}\mbox{.}(2012){Macci{\`o}}, {Paduroiu},
  {Anderhalden}, {Schneider}, \& {Moore}}]{2012MNRAS.424.1105M}
{Macci{\`o}} A.~V., {Paduroiu} S., {Anderhalden} D., {Schneider} A., {Moore}
  B., 2012, \mnras, 424, 1105

\bibitem[{{Maga{\~n}a} \& {Matos}(2012)}]{2012JPhCS.378a2012M}
{Maga{\~n}a} J., {Matos} T., 2012, Journal of Physics Conference Series, 378,
  012012

\bibitem[{{Mashchenko} {et~al}\mbox{.}(2008){Mashchenko}, {Wadsley}, \&
  {Couchman}}]{2008Sci...319..174M}
{Mashchenko} S., {Wadsley} J., {Couchman} H.~M.~P., 2008, Science, 319, 174

\bibitem[{{Mateo}(1998)}]{1998ARA&A..36..435M}
{Mateo} M.~L., 1998, \araa, 36, 435

\bibitem[{{McConnachie}(2012)}]{2012AJ....144....4M}
{McConnachie} A.~W., 2012, \aj, 144, 4

\bibitem[{{McGaugh}(2012)}]{2012AJ....143...40M}
{McGaugh} S.~S., 2012, \aj, 143, 40

\bibitem[{{McGaugh} {et~al}\mbox{.}(2000){McGaugh}, {Schombert}, {Bothun}, \&
  {de Blok}}]{2000ApJ...533L..99M}
{McGaugh} S.~S., {Schombert} J.~M., {Bothun} G.~D., {de Blok} W.~J.~G., 2000,
  \apjl, 533, L99

\bibitem[{{McGaugh} \& {Wolf}(2010)}]{2010ApJ...722..248M}
{McGaugh} S.~S., {Wolf} J., 2010, \apj, 722, 248

\bibitem[{{McQuinn} {et~al}\mbox{.}(2015){McQuinn}, {Lelli}, {Skillman},
  {Dolphin}, {McGaugh}, \& {Williams}}]{2015MNRAS.450.3886M}
{McQuinn} K.~B.~W., {Lelli} F., {Skillman} E.~D., {Dolphin} A.~E., {McGaugh}
  S.~S., {Williams} B.~F., 2015, \mnras, 450, 3886

\bibitem[{{Melotte}(1926)}]{1926MNRAS..86..636M}
{Melotte} P.~J., 1926, \mnras, 86, 636

\bibitem[{{Meyer} {et~al}\mbox{.}(2016){Meyer}, {Meyer}, {Obreschkow}, \&
  {Staveley-Smith}}]{2016MNRAS.455.3136M}
{Meyer} S.~A., {Meyer} M., {Obreschkow} D., {Staveley-Smith} L., 2016, \mnras,
  455, 3136

\bibitem[{{Milgrom}(1983)}]{1983ApJ...270..365M}
{Milgrom} M., 1983, \apj, 270, 365

\bibitem[{{Moffat}(2006)}]{2006JCAP...03..004M}
{Moffat} J.~W., 2006, JCAP, 3, 4

\bibitem[{{Moore}(1994)}]{1994Natur.370..629M}
{Moore} B., 1994, \nat, 370, 629

\bibitem[{{Nakamura} \& {Umemura}(2001)}]{2001ApJ...548...19N}
{Nakamura} F., {Umemura} M., 2001, \apj, 548, 19

\bibitem[{{Navarro} {et~al}\mbox{.}(1996{\natexlab{a}}){Navarro}, {Eke}, \&
  {Frenk}}]{1996MNRAS.283L..72N}
{Navarro} J.~F., {Eke} V.~R., {Frenk} C.~S., 1996{\natexlab{a}}, \mnras, 283,
  L72

\bibitem[{{Navarro} {et~al}\mbox{.}(1996{\natexlab{b}}){Navarro}, {Frenk}, \&
  {White}}]{1996ApJ...462..563N}
{Navarro} J.~F., {Frenk} C.~S., {White} S.~D.~M., 1996{\natexlab{b}}, \apj,
  462, 563

\bibitem[{{Nilson}(1973)}]{1973ugcg.book.....N}
{Nilson} P., 1973, {Uppsala general catalogue of galaxies}

\bibitem[{{O{\~n}orbe} {et~al}\mbox{.}(2015){O{\~n}orbe}, {Boylan-Kolchin},
  {Bullock}, {Hopkins}, {Ker{\v e}s}, {Faucher-Gigu{\`e}re}, {Quataert}, \&
  {Murray}}]{2015arXiv150202036O}
{O{\~n}orbe} J., {Boylan-Kolchin} M., {Bullock} J.~S., {Hopkins} P.~F., {Ker{\v
  e}s} D., {Faucher-Gigu{\`e}re} C.-A., {Quataert} E., {Murray} N., 2015, ArXiv
  e-prints

\bibitem[{{Oh} {et~al}\mbox{.}(2011){Oh}, {Brook}, {Governato}, {Brinks},
  {Mayer}, {de Blok}, {Brooks}, \& {Walter}}]{2011AJ....142...24O}
{Oh} S.-H., {Brook} C., {Governato} F., {Brinks} E., {Mayer} L., {de Blok}
  W.~J.~G., {Brooks} A., {Walter} F., 2011, \aj, 142, 24

\bibitem[{{Oh} {et~al}\mbox{.}(2015){Oh}, {Hunter}, {Brinks}, {Elmegreen},
  {Schruba}, {Walter}, {Rupen}, {Young}, {Simpson}, {Johnson}, {Herrmann},
  {Ficut-Vicas}, {Cigan}, {Heesen}, {Ashley}, \& {Zhang}}]{2015AJ....149..180O}
{Oh} S.-H. {et~al.}, 2015, \aj, 149, 180

\bibitem[{{Oman} {et~al}\mbox{.}(2015){Oman}, {Navarro}, {Fattahi}, {Frenk},
  {Sawala}, {White}, {Bower}, {Crain}, {Furlong}, {Schaller}, {Schaye}, \&
  {Theuns}}]{2015arXiv150401437O}
{Oman} K.~A. {et~al.}, 2015, ArXiv e-prints

\bibitem[{{Oman} {et~al}\mbox{.}(2016){Oman}, {Navarro}, {Sales}, {Fattahi},
  {Frenk}, {Sawala}, {Schaller}, \& {White}}]{2016arXiv160101026O}
{Oman} K.~A., {Navarro} J.~F., {Sales} L.~V., {Fattahi} A., {Frenk} C.~S.,
  {Sawala} T., {Schaller} M., {White} S.~D.~M., 2016, ArXiv e-prints

\bibitem[{{Pe{\~n}arrubia} {et~al}\mbox{.}(2014){Pe{\~n}arrubia}, {Ma},
  {Walker}, \& {McConnachie}}]{2014MNRAS.443.2204P}
{Pe{\~n}arrubia} J., {Ma} Y.-Z., {Walker} M.~G., {McConnachie} A., 2014,
  \mnras, 443, 2204

\bibitem[{{Planck Collaboration} {et~al}\mbox{.}(2013){Planck Collaboration},
  {Ade}, {Aghanim}, {Armitage-Caplan}, {Arnaud}, {Ashdown}, {Atrio-Barandela},
  {Aumont}, {Baccigalupi}, {Banday}, \& et~al.}]{2013arXiv1303.5076P}
{Planck Collaboration} {et~al.}, 2013, ArXiv e-prints

\bibitem[{{Pontzen} \& {Governato}(2012)}]{2012MNRAS.421.3464P}
{Pontzen} A., {Governato} F., 2012, \mnras, 421, 3464

\bibitem[{{Pontzen} \& {Governato}(2014)}]{2014Natur.506..171P}
{Pontzen} A., {Governato} F., 2014, \nat, 506, 171

\bibitem[{{Pontzen} {et~al}\mbox{.}(2015){Pontzen}, {Read}, {Teyssier},
  {Governato}, {Gualandris}, {Roth}, \& {Devriendt}}]{2015arXiv150207356P}
{Pontzen} A., {Read} J., {Teyssier} R., {Governato} F., {Gualandris} A., {Roth}
  N., {Devriendt} J., 2015, ArXiv e-prints

\bibitem[{{Pontzen} {et~al}\mbox{.}(2013){Pontzen}, {Ro{\v s}kar}, {Stinson},
  \& {Woods}}]{2013ascl.soft05002P}
{Pontzen} A., {Ro{\v s}kar} R., {Stinson} G., {Woods} R., 2013, {pynbody:
  N-Body/SPH analysis for python}. Astrophysics Source Code Library

\bibitem[{{Power} {et~al}\mbox{.}(2003){Power}, {Navarro}, {Jenkins}, {Frenk},
  {White}, {Springel}, {Stadel}, \& {Quinn}}]{2003MNRAS.338...14P}
{Power} C., {Navarro} J.~F., {Jenkins} A., {Frenk} C.~S., {White} S.~D.~M.,
  {Springel} V., {Stadel} J., {Quinn} T., 2003, \mnras, 338, 14

\bibitem[{{Read}(2014)}]{2014JPhG...41f3101R}
{Read} J.~I., 2014, Journal of Physics G Nuclear Physics, 41, 063101

\bibitem[{{Read} {et~al}\mbox{.}(2016){Read}, {Agertz}, \&
  {Collins}}]{2015arXiv150804143R}
{Read} J.~I., {Agertz} O., {Collins} M.~L.~M., 2016, \mnras, 459, 2573

\bibitem[{{Read} \& {Gilmore}(2005)}]{2005MNRAS.356..107R}
{Read} J.~I., {Gilmore} G., 2005, \mnras, 356, 107

\bibitem[{{Read} {et~al}\mbox{.}(2006){Read}, {Wilkinson}, {Evans}, {Gilmore},
  \& {Kleyna}}]{2006MNRAS.tmp..153R}
{Read} J.~I., {Wilkinson} M.~I., {Evans} N.~W., {Gilmore} G., {Kleyna} J.~T.,
  2006, \mnras, 367, 387

\bibitem[{{Rhee} {et~al}\mbox{.}(2004){Rhee}, {Valenzuela}, {Klypin},
  {Holtzman}, \& {Moorthy}}]{2004ApJ...617.1059R}
{Rhee} G., {Valenzuela} O., {Klypin} A., {Holtzman} J., {Moorthy} B., 2004,
  \apj, 617, 1059

\bibitem[{{Rubin} \& {Ford}(1970)}]{1970ApJ...159..379R}
{Rubin} V.~C., {Ford}, Jr. W.~K., 1970, \apj, 159, 379

\bibitem[{{Rubin} {et~al}\mbox{.}(1980){Rubin}, {Ford}, \&
  {.~Thonnard}}]{1980ApJ...238..471R}
{Rubin} V.~C., {Ford} W.~K.~J., {.~Thonnard} N., 1980, \apj, 238, 471

\bibitem[{{Ryan-Weber} {et~al}\mbox{.}(2008){Ryan-Weber}, {Begum}, {Oosterloo},
  {Pal}, {Irwin}, {Belokurov}, {Evans}, \& {Zucker}}]{2008MNRAS.384..535R}
{Ryan-Weber} E.~V., {Begum} A., {Oosterloo} T., {Pal} S., {Irwin} M.~J.,
  {Belokurov} V., {Evans} N.~W., {Zucker} D.~B., 2008, \mnras, 384, 535

\bibitem[{{Saitoh} {et~al}\mbox{.}(2008){Saitoh}, {Daisaka}, {Kokubo},
  {Makino}, {Okamoto}, {Tomisaka}, {Wada}, \& {Yoshida}}]{2008PASJ...60..667S}
{Saitoh} T.~R., {Daisaka} H., {Kokubo} E., {Makino} J., {Okamoto} T.,
  {Tomisaka} K., {Wada} K., {Yoshida} N., 2008, \pasj, 60, 667

\bibitem[{{S{\'a}nchez-Salcedo}(2001)}]{2001ApJ...563..867S}
{S{\'a}nchez-Salcedo} F.~J., 2001, \apj, 563, 867

\bibitem[{{Schneider} {et~al}\mbox{.}(1990){Schneider}, {Thuan}, {Magri}, \&
  {Wadiak}}]{1990ApJS...72..245S}
{Schneider} S.~E., {Thuan} T.~X., {Magri} C., {Wadiak} J.~E., 1990, \apjs, 72,
  245

\bibitem[{{Scowcroft} {et~al}\mbox{.}(2013){Scowcroft}, {Freedman}, {Madore},
  {Monson}, {Persson}, {Seibert}, {Rigby}, \&
  {Melbourne}}]{2013ApJ...773..106S}
{Scowcroft} V., {Freedman} W.~L., {Madore} B.~F., {Monson} A.~J., {Persson}
  S.~E., {Seibert} M., {Rigby} J.~R., {Melbourne} J., 2013, \apj, 773, 106

\bibitem[{{Sellwood} \& {S{\'a}nchez}(2010)}]{2010MNRAS.404.1733S}
{Sellwood} J.~A., {S{\'a}nchez} R.~Z., 2010, \mnras, 404, 1733

\bibitem[{{Silich} {et~al}\mbox{.}(2006){Silich}, {Lozinskaya}, {Moiseev},
  {Podorvanuk}, {Rosado}, {Borissova}, \&
  {Valdez-Gutierrez}}]{2006A&A...448..123S}
{Silich} S., {Lozinskaya} T., {Moiseev} A., {Podorvanuk} N., {Rosado} M.,
  {Borissova} J., {Valdez-Gutierrez} M., 2006, \aap, 448, 123

\bibitem[{{Spergel} \& {Steinhardt}(2000)}]{2000PhRvL..84.3760S}
{Spergel} D.~N., {Steinhardt} P.~J., 2000, Physical Review Letters, 84, 3760

\bibitem[{{Strigari} {et~al}\mbox{.}(2007){Strigari}, {Kaplinghat}, \&
  {Bullock}}]{2007PhRvD..75f1303S}
{Strigari} L.~E., {Kaplinghat} M., {Bullock} J.~S., 2007, \prd, 75, 061303

\bibitem[{{Swaters} {et~al}\mbox{.}(2003){Swaters}, {Madore}, {van den Bosch},
  \& {Balcells}}]{2003ApJ...583..732S}
{Swaters} R.~A., {Madore} B.~F., {van den Bosch} F.~C., {Balcells} M., 2003,
  \apj, 583, 732

\bibitem[{{Swaters} {et~al}\mbox{.}(1997){Swaters}, {Sancisi}, \& {van der
  Hulst}}]{1997ApJ...491..140S}
{Swaters} R.~A., {Sancisi} R., {van der Hulst} J.~M., 1997, \apj, 491, 140

\bibitem[{{Tenorio-Tagle} \& {Bodenheimer}(1988)}]{1988ARA&A..26..145T}
{Tenorio-Tagle} G., {Bodenheimer} P., 1988, \araa, 26, 145

\bibitem[{{Tenorio-Tagle} {et~al}\mbox{.}(1987){Tenorio-Tagle}, {Franco},
  {Bodenheimer}, \& {Rozyczka}}]{1987A&A...179..219T}
{Tenorio-Tagle} G., {Franco} J., {Bodenheimer} P., {Rozyczka} M., 1987, \aap,
  179, 219

\bibitem[{{Teyssier}(2002)}]{teyssier02}
{Teyssier} R., 2002, \aap, 385, 337

\bibitem[{{Teyssier} {et~al}\mbox{.}(2013){Teyssier}, {Pontzen}, {Dubois}, \&
  {Read}}]{2013MNRAS.429.3068T}
{Teyssier} R., {Pontzen} A., {Dubois} Y., {Read} J.~I., 2013, \mnras, 429, 3068

\bibitem[{{Truelove} {et~al}\mbox{.}(1997){Truelove}, {Klein}, {McKee},
  {Holliman}, {Howell}, \& {Greenough}}]{1997ApJ...489L.179T}
{Truelove} J.~K., {Klein} R.~I., {McKee} C.~F., {Holliman}, II J.~H., {Howell}
  L.~H., {Greenough} J.~A., 1997, \apjl, 489, L179

\bibitem[{{Trujillo-Gomez} {et~al}\mbox{.}(2015){Trujillo-Gomez}, {Klypin},
  {Col{\'{\i}}n}, {Ceverino}, {Arraki}, \& {Primack}}]{2015MNRAS.446.1140T}
{Trujillo-Gomez} S., {Klypin} A., {Col{\'{\i}}n} P., {Ceverino} D., {Arraki}
  K.~S., {Primack} J., 2015, \mnras, 446, 1140

\bibitem[{{Tully} \& {Fisher}(1977)}]{1977A&A....54..661T}
{Tully} R.~B., {Fisher} J.~R., 1977, \aap, 54, 661

\bibitem[{{Tully} \& {Fisher}(1988)}]{1988cng..book.....T}
{Tully} R.~B., {Fisher} J.~R., 1988, {Catalog of Nearby Galaxies}

\bibitem[{{Valenzuela} {et~al}\mbox{.}(2007){Valenzuela}, {Rhee}, {Klypin},
  {Governato}, {Stinson}, {Quinn}, \& {Wadsley}}]{2007ApJ...657..773V}
{Valenzuela} O., {Rhee} G., {Klypin} A., {Governato} F., {Stinson} G., {Quinn}
  T., {Wadsley} J., 2007, \apj, 657, 773

\bibitem[{{van Albada} {et~al}\mbox{.}(1985){van Albada}, {Bahcall}, {Begeman},
  \& {Sancisi}}]{1985ApJ...295..305V}
{van Albada} T.~S., {Bahcall} J.~N., {Begeman} K., {Sancisi} R., 1985, \apj,
  295, 305

\bibitem[{{van den Bosch} {et~al}\mbox{.}(2000){van den Bosch}, {Robertson},
  {Dalcanton}, \& {de Blok}}]{2000AJ....119.1579V}
{van den Bosch} F.~C., {Robertson} B.~E., {Dalcanton} J.~J., {de Blok}
  W.~J.~G., 2000, \aj, 119, 1579

\bibitem[{{Villaescusa-Navarro} \& {Dalal}(2011)}]{2011JCAP...03..024V}
{Villaescusa-Navarro} F., {Dalal} N., 2011, JCAP, 3, 24

\bibitem[{{Volders}(1959)}]{1959BAN....14..323V}
{Volders} L.~M.~J.~S., 1959, \bain, 14, 323

\bibitem[{{Wada} {et~al}\mbox{.}(2000){Wada}, {Spaans}, \&
  {Kim}}]{2000ApJ...540..797W}
{Wada} K., {Spaans} M., {Kim} S., 2000, \apj, 540, 797

\bibitem[{{Walker} {et~al}\mbox{.}(2009){Walker}, {Mateo}, {Olszewski},
  {Pe{\~n}arrubia}, {Wyn Evans}, \& {Gilmore}}]{2009ApJ...704.1274W}
{Walker} M.~G., {Mateo} M., {Olszewski} E.~W., {Pe{\~n}arrubia} J., {Wyn Evans}
  N., {Gilmore} G., 2009, \apj, 704, 1274

\bibitem[{{Weisz} {et~al}\mbox{.}(2012){Weisz}, {Johnson}, {Johnson},
  {Skillman}, {Lee}, {Kennicutt}, {Calzetti}, {van Zee}, {Bothwell},
  {Dalcanton}, {Dale}, \& {Williams}}]{2012ApJ...744...44W}
{Weisz} D.~R. {et~al.}, 2012, \apj, 744, 44

\bibitem[{{Weldrake} {et~al}\mbox{.}(2003){Weldrake}, {de Blok}, \&
  {Walter}}]{2003MNRAS.340...12W}
{Weldrake} D.~T.~F., {de Blok} W.~J.~G., {Walter} F., 2003, \mnras, 340, 12

\bibitem[{{Wolf} {et~al}\mbox{.}(2010){Wolf}, {Martinez}, {Bullock},
  {Kaplinghat}, {Geha}, {Mu{\~n}oz}, {Simon}, \& {Avedo}}]{2010MNRAS.406.1220W}
{Wolf} J., {Martinez} G.~D., {Bullock} J.~S., {Kaplinghat} M., {Geha} M.,
  {Mu{\~n}oz} R.~R., {Simon} J.~D., {Avedo} F.~F., 2010, \mnras, 406, 1220

\bibitem[{{Wolf}(1909)}]{1909AN....183..187W}
{Wolf} M., 1909, Astronomische Nachrichten, 183, 187

\bibitem[{{Young} {et~al}\mbox{.}(2003){Young}, {van Zee}, {Lo}, {Dohm-Palmer},
  \& {Beierle}}]{2003ApJ...592..111Y}
{Young} L.~M., {van Zee} L., {Lo} K.~Y., {Dohm-Palmer} R.~C., {Beierle} M.~E.,
  2003, \apj, 592, 111

\bibitem[{{Zavala} {et~al}\mbox{.}(2013){Zavala}, {Vogelsberger}, \&
  {Walker}}]{2013MNRAS.431L..20Z}
{Zavala} J., {Vogelsberger} M., {Walker} M.~G., 2013, \mnras, 431, L20

\bibitem[{{Zhang} {et~al}\mbox{.}(2012){Zhang}, {Hunter}, {Elmegreen}, {Gao},
  \& {Schruba}}]{2012AJ....143...47Z}
{Zhang} H.-X., {Hunter} D.~A., {Elmegreen} B.~G., {Gao} Y., {Schruba} A., 2012,
  \aj, 143, 47

\bibitem[{{Zhao} {et~al}\mbox{.}(2006){Zhao}, {Bacon}, {Taylor}, \&
  {Horne}}]{2006MNRAS.368..171Z}
{Zhao} H., {Bacon} D.~J., {Taylor} A.~N., {Horne} K., 2006, \mnras, 368, 171

\end{thebibliography}

\end{document}